\documentclass[11pt,a4paper]{article}

\usepackage{epsfig}
\usepackage{amsmath}
\usepackage{amssymb}
\usepackage{verbatim}
\usepackage{bm}
\usepackage{dsfont}
\usepackage[footnotesize]{caption}
\usepackage{subfigure}
\usepackage{cancel}
\usepackage{cite}
\usepackage{textcomp}
\usepackage{calc}
\usepackage{geometry}
\usepackage{bbm}
\usepackage{color}
\usepackage{hyperref}

\begin{document}


\noindent October 2014 \hfill \parbox{\widthof{SISSA  15/2014/FISI}}{IPMU 14-0319}

\hfill SISSA  15/2014/FISI

\vskip 1.5cm

\begin{center}
{\LARGE\bf Dynamical D-Terms in Supergravity}

\vskip 2cm

{\large Valerie~Domcke$^a$, Kai~Schmitz$^b$, Tsutomu~T.~Yanagida$^b$}\\[3mm]
{\it{
$^a$ SISSA/INFN, 34100 Trieste, Italy \\
$^b$ Kavli IPMU (WPI), University of Tokyo, Kashiwa 277-8583, Japan}}
\end{center}

\vskip 1cm




\begin{abstract}


\noindent Most phenomenological models of supersymmetry breaking
rely on nonzero F-terms rather than nonzero D-terms.
An important reason why D-terms are often neglected is that
it turns out to be very challenging to realize D-terms at energies
parametrically smaller than the Planck scale in supergravity.
As we demonstrate in this paper, all conventional difficulties may, however,
be overcome if the generation of the D-term is based on strong dynamics.
To illustrate our idea, we focus on a certain class of vector-like
SUSY breaking models that enjoy a minimal particle content and which may be easily
embedded into more complete scenarios.
We are then able to show that, upon gauging a global flavor symmetry,
an appropriate choice of Yukawa couplings readily allows to dynamically
generate a D-term at an almost arbitrary energy scale.
This includes in particular the natural and consistent realization of D-terms
around, above and below the scale of grand unification in supergravity,
without the need for fine-tuning of any model parameters.
Our construction might therefore bear the potential to open
up a new direction for model building in supersymmetry and early universe
cosmology.
\end{abstract}


\thispagestyle{empty}

\newpage


\tableofcontents


\section{Introduction and preliminaries}
\label{sec:introduction}


In this paper, we wish to illustrate how an effective Fayet-Iliopoulos (FI) D-term may
be dynamically generated at an intermediate energy scale in strongly interacting
supersymmetric gauge theories.
In Secs.~\ref{subsec:introconstantFI} and \ref{subsec:introeffectiveFI}, we shall
first review the well-known problems related to the existing constructions of constant
and effective FI-terms in the literature.
The reader acquainted with these issues may therefore directly skip to
Sec.~\ref{subsec:introdynamicalFI}, in which we outline our basic idea.


\subsection{Constant field-independent FI-terms in supergravity}
\label{subsec:introconstantFI}


In any realistic supersymmetric extension of the standard model, supersymmetry (SUSY) needs
to be spontaneously broken in some hidden sector.
The order parameters of spontaneous SUSY breaking in a given supersymmetric theory are
the expectation values of the auxiliary $F$ and $D$ fields.
While models that break SUSY via nonzero F-terms are referred to as
O'Raifeartaigh models~\cite{O'Raifeartaigh:1975pr},
models based on nonzero D-terms always feature a realization
of the Fayet-Iliopoulos mechanism~\cite{Fayet:1974jb},
which is why they are also known as FI models of SUSY breaking.
The crucial observation behind SUSY breaking via (Abelian) D-terms is that
the Lagrangian $\mathcal{L}$ of a $U(1)$ gauge theory also admits
the following supersymmetric and gauge-invariant operator,
\begin{align}
\mathcal{L}_{\rm FI} = \int d^4 \theta \, K_{\rm FI} = -g\,\xi \, D
\,, \quad K_{\rm FI} = -2g\, \xi \,V \,,
\label{eq:LagFI}
\end{align}
where $K_{\rm FI}$ is part of the K\"ahler potential,
$V\sim\left(\lambda,A,D\right)$ represents the vector
superfield containing the $U(1)$ gauge degrees of freedom (DOFs),
$\xi$ is a free parameter of mass-dimension $2$,
$g$ stands for the $U(1)$ gauge coupling constant,
and $\theta$ denotes the anticommuting superspace coordinate.
If one manages to stabilize all scalars carrying nonzero $U(1)$ gauge charge around their
origin, the operator in Eq.~\eqref{eq:LagFI} leads to a nonvanishing D-term scalar
potential belonging to the $U(1)$ gauge interactions, $V_D \propto g^2\xi^2$, and
hence to the spontaneous breaking of SUSY.
This mechanism has several interesting phenomenological applications in
supersymmetric model building as well as in cosmology.
A nonvanishing FI-term $\mathcal{L}_{\rm FI}$ can, for instance, play a crucial
role in mediating SUSY breaking to the visible sector
or provide the vacuum energy density that is necessary to drive the
inflationary stage in the very early universe~\cite{Binetruy:1996xj}.


Despite their abundant occurrence in the literature on SUSY phenomenology
over the last four decades, some important aspects of FI-terms have, however,
become clear only in the past few
years~\cite{Komargodski:2009pc,Dienes:2009td,Komargodski:2010rb}.
As it turns out, it is in fact very difficult or even impossible to consistently
couple a $U(1)$ gauge theory featuring a genuine (i.e.\ constant, field-independent)
FI-term to minimal supergravity (SUGRA).
If the FI parameter $\xi$ is assumed to be a fundamental constant, coupling
the rigid theory to gravity requires that the final locally supersymmetric
theory must have an additional exact global continuous symmetry.%
\footnote{This conclusion can be avoided if the rigid theory only contains fields
with vanishing $U(1)$ charge, cf.\ Ref.~\cite{Catino:2011mu} for an explicit
model, or if $\xi/2$ is quantized in units of the reduced Planck
mass $M_{\rm Pl} = \left(8\pi G\right)^{-1/2}$~\cite{Witten:1982hu}.
The latter is always the case once the underlying $U(1)$ gauge group is assumed
to be compact, i.e.\ when its global topology is that of a true $U(1)$
and not the one of the real numbers $\mathbb{R}$.
For noncompact global topology, $\xi$ can be parametrically small,
$\xi/M_{\rm Pl}^2 \ll 1$; but then the SUGRA theory needs to exhibit a
global continuous symmetry.\smallskip}
As shown in Refs.~\cite{Komargodski:2009pc,Komargodski:2010rb}, this result is
independent of the SUGRA formalism and equally
applies in the old~\cite{Stelle:1978ye} as well as in the new
minimal~\cite{Akulov:1976ck} off-shell formulation of SUGRA.
According to general rules of quantum gravity, all global symmetries are, however,
eventually broken by gravity effects~\cite{Giddings:1988cx}.%
\footnote{Even if one disregards this conceptional argument about the general
properties of quantum gravity, theories with a constant FI-term are still in
trouble for phenomenological reasons.
Upon the coupling to SUGRA, the initial non-$R$ $U(1)$ gauge symmetry
turns into a continuous local $R$ symmetry.
In the context of the standard model, every such symmetry is, however, necessarily
anomalous, which renders the entire theory inconsistent at the quantum level.
We are thankful to W.\ Buchm\"uller and R.\ Kallosh for a helpful discussion
on this point.}
A theory based on minimal SUGRA and exhibiting a global symmetry is
therefore necessarily inconsistent.
This also explains why all attempts to find string compactifications
with genuine FI-terms in the low-energy effective theory have been futile so far.


\subsection{Effective field-dependent FI-terms from string theory}
\label{subsec:introeffectiveFI}


A possible way out of these difficulties is to resort to field-dependent FI-terms,
in the case of which $\xi$ is regarded as an effective parameter that actually
depends on the vacuum expectation values (VEVs) of other scalar
fields, $\xi = \xi\left(\left<\phi_i\right>\right)$.
The generation of such field-dependent FI-terms is therefore always associated
with the spontaneous breaking of the $U(1)$ gauge symmetry and one should actually
not refer to them as FI-terms.
Instead, they merely correspond to the VEV of the auxiliary $D$ field in the new vacuum
after spontaneous symmetry breaking, $\xi \equiv \left<D\right>/g$.
As pointed in Ref.~\cite{Komargodski:2009pc}, the fundamental obstacle in coupling a
theory with a constant FI-term to SUGRA, which eventually also necessitates the introduction
of a global symmetry, is the fact that such theories do not possess
a gauge-invariant Ferrara-Zumino (FZ) supercurrent multiplet~\cite{Ferrara:1974pz}.
By appropriately choosing the gauge transformation behavior of the fields
$\phi_i$, the gauge invariance of the FZ-multiplet can, however, be preserved
and the theory can be consistently coupled to SUGRA within the old minimal formalism.
A famous example of such a construction are the field-dependent FI-terms
frequently encountered in string theory~\cite{Dine:1987xk}, which are based on the
Green-Schwarz mechanism~\cite{Green:1984sg} of anomaly cancellation,
\begin{align}
K_{\rm FI} =  f_{\rm GS}\big(V + \Phi + \Phi^\dagger\big) \,.
\label{eq:KFIstring}
\end{align}
Here, $f_{\rm GS}$ is an appropriate function of the linear combination $V + \Phi + \Phi^\dagger$
and $\Phi$ stands for a (not necessarily properly normalized) modulus field that transforms
in the affine representation of the (noncompact) $U(1)$ gauge group.
An alternative approach to deal with the non-gauge invariance of the FZ-multiplet
in the presence of a constant FI-term is to trade the FZ-multiplet for the
so-called $\mathcal{S}$-multiplet~\cite{Komargodski:2010rb}, which is always well-defined.
Gauging the $\mathcal{S}$-multiplet rather than the FZ-multiplet then amounts to
coupling the rigid theory to $16/16$ SUGRA~\cite{Girardi:1984vq} rather than
to minimal SUGRA.
In this non-minimal framework for SUGRA, the gravity multiplet contains
an additional chiral matter multiplet next to the ordinary graviton and the
ordinary gravitino.
Interestingly enough, this additional chiral field can be identified with the
above modulus field $\Phi$ and the FI-term in the gauged theory ends up being
of the same form as in Eq.~\eqref{eq:KFIstring}.


Now one, however, faces the problem that the modulus $\Phi$ needs to be
stabilized at sufficiently high energies, since it would otherwise
absorb the effective FI-term in its VEV.
This requirement imposes strong constraints on the underlying high-energy theory,
which may be hard to fulfill.
One possibility in this context could potentially be to rely on a large
gravity-mediated mass $m_\phi$ for the modulus field.
Depending on the size of the effective FI parameter $\xi_{\rm GS}$,
this would, however, require an extremely large gravitino mass,
$m_\phi \sim m_{3/2} \gtrsim g\sqrt{\left|\xi_{\rm GS}\right|}$~\cite{Wieck:2014xxa}.
Alternatively, one may attempt to stabilize the modulus above the
SUSY breaking scale by means of a dedicated mechanism.
In this case, the vector multiplet $V$ will, however, acquire the same mass as
the modulus $\Phi$ via the St\"uckelberg mechanism.
Then, once we integrate out the modulus at low energies, also the vector multiplet
decouples, such that there is no energy range in which we could meaningfully speak of an
effective FI-term for the $U(1)$ vector field.
Besides this, even more elaborate attempts to stabilize $\Phi$ are not guaranteed
to be successful.
In Ref.~\cite{Wieck:2014xxa}, it has, for instance, recently been shown that,
in the context of ordinary D-term hybrid inflation, all of the standard,
straightforward approaches to stabilize the modulus field $\Phi$ are bound to fail.


\subsection{Dynamical FI-terms in strongly interacting gauge theories}
\label{subsec:introdynamicalFI}


Because of these limitations of the existing constructions of effective FI-terms
in the literature, it is desirable to seek alternative mechanisms for the generation
of field-dependent FI-terms which are consistent also in the presence of gravity
and which, at the same time, do not lead to any problems related to the modulus field $\Phi$.
An attractive possibility in this context, which we will further explore
in this paper, is to base the generation of the FI-term on the dynamics of
strongly interacting supersymmetric gauge theories.%
\footnote{The first model exploiting this possibility to generate a dynamical FI-term
has been presented in Ref.~\cite{Nakayama:2007du}.}
Here, our main observation is that, in models of dynamical SUSY breaking (DSB),
it is possible to generate nonvanishing D-terms by gauging a global $U(1)$ flavor
symmetry, followed by adjusting the VEVs of the resulting charged composite fields
at low energies by means of appropriate Yukawa interactions.


In this sense, our construction bears some resemblance to the mechanism
described in Ref.~\cite{Dumitrescu:2010ca}, which also utilizes
strong dynamics to generate an effective D-term.
Instead of a superpotential suited for dynamical SUSY breaking,
this mechanism, however, relies on a runaway superpotential.
While our F-term scalar potential exhibits a stable SUSY-breaking vacuum
from the very beginning, the corresponding scalar potential analyzed
in Ref.~\cite{Dumitrescu:2010ca} therefore initially comes with a
supersymmetric vacuum at infinity.
After weakly gauging a global symmetry (just as in our case), 
the runaway directions in the F-term potential are then
stabilized by the D-term contributions to the scalar potential.
A further crucial difference between our mechanism and the
one presented in Ref.~\cite{Dumitrescu:2010ca} is that we
focus on simple vector-like gauge theories, while Ref.~\cite{Dumitrescu:2010ca}
only discusses a set of chiral models.
Our mechanism hence appears to be more minimal and promises to
be more easily applicable in the explicit construction of realistic models.


Now, to see how an effective D-term may be generated in a given vector-like
DSB model, imagine that the low-energy DOFs of this theory correspond to, for
instance, a set of mesons $M^i$.
Further, suppose that the low-energy effective theory contains a global $U(1)$
flavor symmetry, under which the meson fields carry charges $q_i$.
We are then free to gauge this flavor symmetry, which provides us with a $U(1)$ D-term
of the following form,
\begin{align}
D = - g \sum_i q_i \left|M^i\right|^2 + \Delta D \,, \quad
\sum_{\rm all} q = \sum_{\rm all} q^3 = 0 \,,
\label{eq:DMsum}
\end{align}
where $\Delta D$ stands for further contributions to $D$ from additional
charged fields and where we implicitly assume that at least some of the
charges $q_i$ are nonzero.
Also, note that the sum of all charges as well as the sum of all charges
cubed are required to vanish in order to ensure anomaly-freedom.
As an elementary ingredient of our construction, we emphasize that,
in the context of dynamical SUSY breaking,
all flat directions in moduli space are necessarily lifted.
The mesons are thus guaranteed to acquire well-defined and definite VEVs,
\begin{align}
\left<M^i\right> = f_i \left(\lambda_j,g\right) \Lambda \,.
\label{eq:Mvev}
\end{align}
Here, $\Lambda$ is the dynamical scale of the strong interactions and the $f_i$
are model-dependent functions of the Yukawa coupling constants $\lambda_j$ in the theory
as well as of the gauge coupling constant $g$.
Note that, in contrast to the corresponding scalar VEVs discussed in
Ref.~\cite{Dumitrescu:2010ca}, our meson VEVs are also still well-behaved
in the limit $g\rightarrow 0$.
For appropriate functions $f_i$, it is then straightforward to generate
a nonzero D-term proportional to the dynamical scale, $D \sim g\,\Lambda^2$.


By construction, the such obtained dynamical D-terms can never be the only source of
SUSY breaking.
Instead, SUSY is always also broken by the strong dynamics responsible for the VEVs
of the composite fields at low energies, cf.\ Eq.~\eqref{eq:Mvev}.
This source of SUSY breaking is associated with one or several nonzero F-terms,
the magnitude of which exceeds the one of the D-term.
This result is consistent with general theorems in SUGRA, which state
that generically the dominant contribution to SUSY breaking is provided by
F-terms rather than by D-terms~\cite{Nakayama:2007du,Dumitrescu:2010ca,Kawamura:2010mb},
\begin{align}
\left|D\right| \lesssim \left|F\right| \,.
\label{eq:DFbound}
\end{align}
An obvious advantage of relying on strong dynamics in generating an effective D-term
is that the magnitude of the such obtained D-term is controlled by the dynamical
scale $\Lambda$, so that it can be easily varied over many orders of magnitude,
\begin{align}
\sqrt{\left|\xi\right|} \sim \Lambda \,, \quad
\Lambda_{\rm min} \lesssim \Lambda \lesssim M_{\rm Pl} \,.
\label{eq:Lambdarange}
\end{align}
Here, $\Lambda_{\rm min}$ denotes a model-dependent phenomenological lower bound
on the dynamical scale, while the Planck mass $M_{\rm Pl}$ represents a
model-independent theoretical upper bound.
Depending on the details of the coupling between the strongly interacting and
the visible sector, we expect $\Lambda_{\rm min}$ to be typically of
$\mathcal{O}\left(100\right)\rm TeV$.
Meanwhile, we point out that values of the dynamical scale exceeding the
Planck scale would take us out of the validity range of SUGRA as a
low-energy effective description of quantum gravity and hence such
large $\Lambda$ values are not admissible.
We also mention that, in the context of a grand unified theory (GUT),
a dynamically generated D-term could very well be of the order
of the unification scale, $\Lambda_{\rm GUT} \simeq 2 \times 10^{16}\,\rm GeV$.
This would certainly be particularly appealing from the perspective
of both particle physics and cosmology.


The range of viable $\Lambda$ values in our dynamical setup, cf.\ Eq.~\eqref{eq:Lambdarange},
needs to be contrasted with the expected size of the anomalous FI-term in string theory.
All relevant energy scales in string theory, the compactification scale $M_c$,
the string scale $M_s$ as well as the four-dimensional Planck scale $M_{\rm Pl}$, are
all very large, $M_c \sim M_s \sim M_{\rm Pl} \sim 10^{18}\,\rm GeV$, which is why, purely
based on dimensional analysis, we would also expect a stringy FI-term to be
very large.
To make this argument a bit more explicit, suppose that the function $f_{\rm GS}$
in Eq.~\eqref{eq:KFIstring} can be expanded as a Taylor series in
$V_M = V + \Phi + \Phi^\dagger$, so that in the vicinity of $V_M=0$ we are able to write%
\footnote{Note that, in order to realize a non-spurious FI-term along with a kinetic
term for the modulus field $\Phi$, this series needs to extend at least up to cubic
order in $V_M$.
Otherwise, the linear term, which actually induces the nonvanishing effective D-term,
could always be shifted away by a field redefinition.}
\begin{align}
f_{\rm GS}\left(V_M\right) \sim M_{\rm Pl}^2 \sum_{n=0}^\infty
\frac{1}{n!}\, c_n V_M^n \,, 
\end{align}
Without any particular fine-tuning among the order $\mathcal{O}(1)$ coefficients $c_n$,
the function $f_{\rm GS}$ is then guaranteed to yield an effective FI-term scale
close to the Planck scale, $\sqrt{\left|\xi_{\rm GS}\right|} \sim M_{\rm Pl}$.
The dynamical generation of an effective D-term in a strongly coupled field theory
is hence superior to its stringy alternative based on the Green-Schwarz mechanism
in the sense that the former is capable of realizing $\xi$ values
in a much larger range than the latter.


In the present paper, we will restrict ourselves to the arguably simplest case and
illustrate our idea only by means of DSB models based on $SP(N_c)$ gauge
dynamics, i.e.\ dynamical models breaking SUSY \`a la IYIT~\cite{Izawa:1996pk}.
To this end, we first describe in detail the minimal case
of an $SP(1)$ theory%
\footnote{In the convention used here, the strongly coupled $SP(1)$ theory
is equivalent to an $SU(2)$ gauge theory.}
in the following section, before we then
comment on the general $SP(N_c)$ case in Sec.~\ref{sec:generalizations}.
Besides this, we also explain in Sec.~\ref{sec:generalizations} why DSB models
based on $SU(N_c)$ instead of $SP(N_c)$ dynamics fail to provide a basis
for the successful generation of an effective D-term.
In Sec.~\ref{sec:applications}, we then sketch what kind of effects a dynamically
generated D-term may have on the superparticle mass spectrum in the visible
sector as well as how it may be used for the construction of inflationary
models. 
Finally, we summarize our results and give an outlook as to how
our study may be continued in Sec.~\ref{sec:conclusions}.


\section{Minimal setup based on \texorpdfstring{\boldmath{$SP(1)$}}{SP(1)} dynamics
(\texorpdfstring{$\cong SU(2)$}{= SU(2)} dynamics)}
\label{sec:minimal}


Our dynamical generation of an effective FI-term will be based on the
IYIT model.
In Sec.~\ref{subsec:IYIT}, we shall first review this model and
outline how it accomplishes the dynamical breaking of SUSY via
the O'Raifeartaigh mechanism.
The reader familiar with this model may directly proceed with
Sec.~\ref{subsec:effectiveFI}, in which we explicitly present our
construction of the field-dependent FI-term.


\subsection{The IYIT model of dynamical supersymmetry breaking}
\label{subsec:IYIT}


The vector-like model introduced in Ref.~\cite{Izawa:1996pk}, sometimes referred to
as the IYIT model, represents a minimal example of a supersymmetric gauge theory
accomplishing spontaneous SUSY breaking by means of strong dynamics.
This model is based on strongly interacting $SP(N_c)$ gauge dynamics
and features $2N_f = 2(N_c + 1)$ chiral quark (i.e.\ matter) fields $Q^i$
transforming in the fundamental representation of $SP(N_c)$.
At energies below the dynamical scale $\Lambda$, the interaction between
these quark fields is best described in terms of the $2 N_f (2 N_f -1)/2$
gauge-invariant composite meson fields $M^{ij} = - M^{ji} = Q^i Q^j/\Lambda$.
Compared to other DSB models, the field content of the low-energy effective
theory is hence rather minimal.
Unlike, for instance, the DSB models based on $SU(N_c)$ dynamics, it only contains
meson fields and no other, more complicated composite states such
as baryons and antibaryons.
A further virtue of the IYIT model is its vector-like matter content,
which facilitates its analysis and which makes it easier to
embed it into more complete scenarios.
Chiral models, such as those presented in Ref.~\cite{Dumitrescu:2010ca},
tend, by contrast, to be more involved and are perhaps less suited for
further generalizations \cite{Schmitz:2014xx}.


For our special choice of quark flavors, $N_f = N_c + 1$, no dynamical
(ADS) superpotential~\cite{Affleck:1983mk} is generated at low energies.
The quantum moduli space is instead simply spanned by
the $N_f(2N_f-1)$ flat meson directions $M^{ij}$,
subject to the following constraint~\cite{Seiberg:1994bz},
\begin{align}
\text{Pf} \left(M^{ij}\right) = \Lambda^{N_c+1} \,,
\label{eq_pfaff}
\end{align}
where $\text{Pf}\left(M\right)$ denotes the Pfaffian of the antisymmetric
meson matrix $M$, $\left[\text{Pf}\left(M\right)\right]^2 = \text{det}\left(M\right)$.
This constraint is the quantum mechanically deformed version of the classical
moduli constraint, $\text{Pf} \left(M\right) = 0$, where the appearance
of the dynamical scale on the right-hand side of Eq.~\eqref{eq_pfaff} is
due to nonperturbative instanton effects.
A convenient way to implement the deformed moduli constraint when
studying the quantum moduli space of the $SP(N_c)$ theory is to include
it directly into the effective superpotential in the form of a Lagrange
constraint term,
\begin{align}
W_{\rm eff} \propto \frac{T}{\Lambda^{N_c-1}}
\left[\text{Pf} \left(M^{ij}\right) - \Lambda^{N_c+1}\right] \,,
\label{eq:WT}
\end{align}
with the chiral superfield $T$ representing a Lagrange multiplier.
The overall normalization of this effective superpotential is
unfortunately uncalculable as the nature of the K\"ahler potential
for the field $T$ is unknown.
$T$ certainly does not possess a perturbative K\"ahler potential
and whether or not it possesses a nonperturbative K\"ahler potential is an
open question.
If strong-coupling effects below the dynamical scale should happen to
generate a K\"ahler potential for $T$, the superpotential in Eq.~\eqref{eq:WT}
would end up having a definite normalization,
\begin{align}
W_{\rm eff} = \lambda_T \,\frac{T}{\Lambda^{N_c-1}}
\left[\text{Pf} \left(M^{ij}\right) - \Lambda^{N_c+1}\right] \,,
\label{eq:WTdef}
\end{align}
with $T$ being canonically normalized and for some finite coupling constant $\lambda_T$. 
If, on the other hand, no K\"ahler potential should be generated,
we would have to interpret $T$ as a mere auxiliary field.
This would then correspond to the limit $\lambda_T \rightarrow \infty$ in
the above superpotential.
As we are unable to calculate the K\"ahler potential for the field $T$,
we will simply decouple all effects related to it in the following.
Practically speaking, we will do so by assuming that the dimensionless
parameter $\lambda_T$ is much larger than all other coupling constants in the theory.


In the above outlined setup, dynamical SUSY breaking is now achieved by
stabilizing all flat directions in moduli space by means of appropriate Yukawa
interactions.
For every flat direction $M^{ij}$, we introduce a chiral singlet field $Z_{ij}$,
which we then couple to the fundamental quark fields in the tree-level superpotential
as follows,%
\footnote{Throughout this paper, the flavor indices $i,j,k,l$ always run from $1$ to $2N_f$.}
\begin{align}
W_{\rm tree} = \frac{1}{2}\,\lambda_{kl}^{ij} \, Z_{ij} \, Q^k Q^l \,, \quad
\lambda_{ij}^{kl} = - \lambda_{ji}^{kl} = - \lambda_{ij}^{lk} \,, \quad Z_{ij} = - Z_{ji} \,.
\label{eq:Wtree}
\end{align}
Here, we assume all complex phases of the $\mathcal{O}(1)$ Yukawa coupling constants
to be absorbed in the singlet fields $Z_{ij}$ for simplicity.
Note that the maximal flavor symmetry of this tree-level superpotential
corresponds to a global $SU\left(2N_f\right)$ symmetry, provided that
all Yukawa couplings are equal, $\lambda_{kl}^{ij} \equiv \lambda$.
We shall, however, only be interested in Abelian subgroups of this maximal
flavor symmetry, which is why we are free to redefine the fields $Z_{ij}$
such that Eq.~\eqref{eq:Wtree} turns into%
\begin{align}
W_{\rm tree} = \frac{1}{2}\,\lambda_{ij} \, Z_{ij} \, Q^i Q^j \,.
\label{eq:WtreeU1}
\end{align}
At energies below the dynamical scale, this superpotential can be reformulated
in terms of the meson fields $M^{ij}$, so that the full effective superpotential
is eventually given as
\begin{align}
W_{\rm eff} \simeq  \lambda_T \,\frac{T}{\Lambda^{N_c-1}}
\left[\text{Pf} \left(M^{ij}\right) - \Lambda^{N_c+1}\right] +
\frac{1}{2}\,\lambda_{ij}\, \Lambda \, Z_{ij} \, M^{ij} \,,
\label{eq:Weff}
\end{align}
where we have neglected all corrections to the Yukawa couplings $\lambda_{ij}$
that arise when running down from high to low energies.
This means in particular that the meson fields in Eq.~\eqref{eq:Weff}
are supposed to represent the canonically normalized DOFs at low energies.
The crucial property of the superpotential in Eq.~\eqref{eq:Weff} is that it
leads to F-term conditions that cannot all be satisfied simultaneously,
as long as none of the Yukawa couplings $\lambda_{ij}$ is actually zero.
SUSY is therefore spontaneously broken via the O'Raifeartaigh mechanism.
On the other hand, in the case of one of the couplings $\lambda_{ij}$ being
zero, the low-energy vacuum is located at infinity in moduli space and
SUSY remains preserved.
Likewise, for more than one Yukawa coupling being zero, we recur to the
original situation, in which the moduli space exhibits a number
of flat directions, along which SUSY is unbroken.
In the following, we will disregard these possibilities and focus on the case
of generic, nonzero Yukawa couplings, so that SUSY is always dynamically broken. 


\subsection[Effective FI-term upon weakly gauging a global \texorpdfstring{$U(1)$}{U(1)} flavor symmetry]
{Effective FI-term upon weakly gauging a global \texorpdfstring{\boldmath{$U(1)$}}{U(1)} flavor symmetry}
\label{subsec:effectiveFI}


In the remainder of this section, we will now focus on the case of $SP(1) \cong SU(2)$
dynamics in combination with $N_f = 2$ quark flavors and illustrate how the IYIT model may
allow for the dynamical generation of an effective FI-term.
In doing so, we will also discuss the magnitude of the SUSY breaking scale
and outline how the fundamental DOFs eventually end up being distributed in
the low-energy effective theory.
Here, we will in particular observe that the role played by some of the
fundamental DOFs turns out to be slightly different than in the usual IYIT
model without an Abelian FI-term.
To start with, let us inspect once more the tree-level superpotential in
Eq.~\eqref{eq:Wtree} for the special case of $N_f = 2$.
This superpotential exhibits an axial $U(1)$ symmetry associated with
a $Q^i$ phase rotation.
It is this $U(1)$ symmetry for which we are now going to generate a
nonvanishing effective D-term.
In the first step, we first of all need to (weakly) gauge this symmetry
and assign appropriate gauge charges to the chiral fields of our model. 
We assign $U(1)$ charges to the quark fields $Q^i$ as follows,
\begin{align}
[Q_1] = [Q_2] = + \frac{1}{2} \,, \quad [Q_3] = [Q_4] = - \frac{1}{2} \,.
\label{eq:Qcharges}
\end{align}
Correspondingly, the six singlet fields $Z_{ij}$ then carry the following charges,
\begin{align}
[Z_{12}] = -1 \,, \quad [Z_{34}] = +1 \,, \quad 
[Z_{13}] = [Z_{14}] = [Z_{23}] = [Z_{24}] = 0 \,.
\label{eq:Zcharges}
\end{align}
For the ease of notation, we will therefore refer to $Z_{12}$ as $Z_-$,
to $Z_{34}$ as $Z_+$ and to $Z_{13}$, $Z_{14}$, $Z_{23}$, and $Z_{24}$
as $Z_1^0$, $Z_2^0$, $Z_3^0$, and $Z_4^0$ in the following.
We emphasize that our charge assignment in Eqs.~\eqref{eq:Qcharges} and \eqref{eq:Zcharges}
is such that $\sum_i q_i = \sum_i q_i^3 = 0$, as required so as
to render the $U(1)$ flavor symmetry anomaly-free, cf.\ Eq.~\eqref{eq:DMsum}.
Furthermore, also the mesons in the low-energy
effective theory now carry $U(1)$ gauge charges.
According to their quark content, the six mesons $M^{ij}$ are charged as follows,
\begin{align}
[M^{12}] = +1 \,, \quad [M^{34}] = -1 \,, \quad 
[M^{13}] = [M^{14}] = [M^{23}] = [M^{24}] = 0 \,.
\end{align}
Similarly as in the case of the singlet fields, we will from now on refer to
$M^{12}$ as $M_+$, to $M^{34}$ as $M_-$ and to $M^{13}$, $M^{14}$, $M^{23}$, and $M^{24}$
as $M_1^0$, $M_2^0$, $M_3^0$, and $M_4^0$.
For an overview of our assignment of $U(1)$ gauge charges, cf.\ also Tab.~\ref{tab_assignment}.


\begin{table}
\centering
\begin{tabular}{l||ccccccccccc} 
Field         &  $Q^1$  & $Q^2$  & $Q^3$  & $Q^4$  & $M_+$ & $M_-$ & $M^0_{a}$ & $Z_+$ & $Z_-$ & $Z^0_{a}$ \\\hline
$U(1)$ charge &  $+1/2$ & $+1/2$ & $-1/2$ & $-1/2$ & $+1$  & $-1$  & $0$       & $+1$  & $-1$  & $0$ \\
\end{tabular}
\caption{Chiral fields present in the minimal $SP(1)$ model (in the high-energy as well as in the
low-energy regime) and charge assignment under the weakly gauged $U(1)$ flavor symmetry.
Here, $a = {1,2,3,4}$.
At low energies, the quark fields form mesons, of which in particular
the charged mesons $M_+ = Q^1 Q^2 / \Lambda$ and $M_- = Q^3 Q^4 / \Lambda$
play an important role in the generation of the effective FI-term.
All of the above charges could in principle also be rescaled, $q \rightarrow nq$,
as long as the gauge coupling constant $g$ is appropriately rescaled, too,
$g \rightarrow g/n$.}
\label{tab_assignment}
\end{table}


In terms of the charge eigenstates at low energies, the effective
superpotential in Eq.~\eqref{eq:Weff} can be rewritten as,
\begin{align}
W_{\rm eff} \simeq  \lambda_T \,T
\left[\text{Pf} \left(M^{ij}\right) - \Lambda^2\right] +
\lambda_+ \Lambda \, M_+ \, Z_- + 
\lambda_- \Lambda \, M_- \, Z_+ +
\lambda_a^0 \Lambda \, M_a^0 \, Z_a^0
\,,\label{eq:Weffcharg}
\end{align}
where $a=1,2,3,4$ and where we have renamed the Yukawa couplings
$\lambda_{ij}$ in Eq.~\eqref{eq:Weff} in an obvious way.
Meanwhile, as we are dealing with the particular case of six mesons,
the Pfaffian $\text{Pf} \left(M\right)$ of the antisymmetric meson matrix
can be readily expanded in the following fashion,
\begin{align}
\text{Pf} \left(M^{ij}\right) = M_+ M_- - M_1^0 M_4^0 + M_2^0 M_3^0 \,.
\label{eq:pfaff}
\end{align}
Eq.~\eqref{eq:Weffcharg} in combination with Eq.~\eqref{eq:pfaff}
allows to compute the F-term scalar potential for the scalar components
of our meson and singlet fields.
As we have gauged the global $U(1)$ flavor symmetry of the tree-level
superpotential in Eq.~\eqref{eq:Wtree}, this potential now needs
to be supplemented by a D-term scalar potential accounting for the
$U(1)$ gauge interactions in the scalar sector, 
\begin{align}
V_D =  \frac{g^2}{2} \left( |M_+|^2 - |M_-|^2 + |Z_+|^2 - |Z_-|^2 + \dots \right)^2 \,,
\label{eq:VD}
\end{align}
where the ellipsis stands for hypothetical particles from other sectors
which also carry $U(1)$ gauge charge.
The central idea behind our mechanism for the dynamical generation of an
effective D-term is now the following:
By appropriately choosing the Yukawa couplings in Eq.~\eqref{eq:Weffcharg}
we are able to engineer the VEVs of the charged fields contributing to $V_D$
in Eq.~\eqref{eq:VD} in such a way that $\left<Z_\pm\right> = 0$ and
$\left<M_+\right> \neq \left<M_-\right>$.
This then results in a nonvanishing effective FI parameter%
\footnote{Unless stated otherwise, we shall always assume that except for
$M_+$ and $M_-$ no other charged field acquires a (large) VEV contributing
to $\xi$.
This means in particular that we shall take it for granted that the effective
$\xi$ parameter in Eq.~\eqref{eq:xiMM} is not inadvertently absorbed by
the VEV of another hidden-sector field.}
\begin{align}
\xi = \big< \left|M_-\right|^2 \big> - \big< \left|M_+\right|^2 \big> \neq 0 \,.
\label{eq:xiMM}
\end{align}


Let us now calculate the VEVs of all scalar meson and singlet fields in our
$SP(1)$ model.
To facilitate our analysis, we assume a (slight) hierarchy between
the Yukawa couplings for the uncharged fields and those for the charged fields,
$\lambda_a^0 \gg \lambda_\pm$.
This automatically guarantees that all neutral fields are stabilized around
the origin, $\left<M_a^0\right> = \left<Z_a^0\right> = 0$.
In consequence of that, the deformed moduli constraint turns into a condition
for the charged mesons $M_+$ and $M_-$ only,
\begin{align}
\text{Pf} \left(M^{ij}\right) = \Lambda^2 \quad\rightarrow\quad
M_+ M_- = \Lambda^2 \,.
\end{align}
This constraint is invariant under $U(1)$ super-gauge transformations,
$M_\pm \rightarrow M_\pm e^{\pm S}$, for some superfield-valued
super-gauge transformation parameter $S$.
A convenient way to parametrize the fluctuations of $M_+$ and $M_-$
around their respective VEVs is hence the following,
\begin{align}
M_+ = \left(\left<M_+\right> + M \right) e^S \,, \quad
M_- = \left(\left<M_-\right> + M \right) e^{-S} \,,
\label{eq:MMMS}
\end{align}
where $M$ and $S$ are chiral superfields of mass dimension $1$ and $0$, respectively.
The advantage of this parametrization is that, later on, it will allow us
to explicitly identify the super-gauge transformation parameter
$S$ with the Goldstone multiplet, which is absorbed by the $U(1)$
vector multiplet $V$ upon the spontaneous breaking of the
$U(1)$ symmetry, cf.\ Sec.~\ref{subsec:spectrum}.
Plugging the expressions in Eq.~\eqref{eq:MMMS} into the effective
superpotential in Eq.~\eqref{eq:Weffcharg}
and setting all neutral meson and singlet fields to $0$, we find
\begin{align}
W_{\rm eff} \simeq & \: \lambda_T \,T
\left[\left<M_+\right> \left<M_-\right>- \Lambda^2\right] +
\lambda_T \left(\left<M_+\right> +  \left<M_-\right>\right) T M +
\lambda_T \, T M^2
\\\nonumber
+ & \: \lambda_+ \Lambda \left(\left<M_+\right> + M\right) e^S \,  Z_- + 
\lambda_- \Lambda \left(\left<M_-\right> + M\right) e^{-S} \, Z_+ \,.
\end{align}
This form of the effective superpotential makes several important
things immediately obvious:
(i) The fields $Z_+$ and $Z_-$ have both nonvanishing F-terms.
They hence both contribute to the goldstino multiplet $X$ responsible for
the spontaneous breaking of SUSY.
(ii) If the field $T$ is indeed dynamical, i.e.\ if $\lambda_T \sim \mathcal{O}(1)$,
it, too, possesses a nonvanishing F-term, rendering it also part of the
goldstino multiplet.
Here, an explicit calculation leads to $F_T \simeq \lambda_+\lambda_-/\lambda_T \,\Lambda$.
If, on the other hand, $T$ is a mere auxiliary field, i.e.\ if
$\lambda_T \rightarrow \infty$, its F-term vanishes
and the deformed moduli constraint ends up being exactly fulfilled
by the VEVs of the charged meson fields,
\begin{align}
\lambda_T \rightarrow \infty  \,, \quad
F_T \rightarrow 0 \,, \quad
\left<M_+\right> \left<M_-\right> - \Lambda^2 \rightarrow 0 \,.
\end{align}
(iii) The fields $T$ and $M$ share a supersymmetric mass term.
Once we require that the deformed moduli constraint be satisfied
exactly, i.e.\ once we send $\lambda_T$ to infinity, this mass
blows up.
The fields $T$ and $M$ thus become very heavy, which causes them to
decouple from the low-energy dynamics.
In this limit, the deformed moduli constraint then eliminates
the (auxiliary) field $T$ as well as the field $M$, i.e.\ one
complete chiral multiplet of mesonic DOFs.
As already mentioned below Eq.~\eqref{eq:WTdef}, we shall work in exactly
this limit in the following.
We reiterate once more that this amounts to considering the field $T$
as a mere undynamical Lagrange multiplier.%
\footnote{From now on, we will therefore simply set $F_T = 0$.
This certainly does not limit the validity of our construction,
because, even in the case of a dynamical field $T$, the F-term $F_T$
is typically subdominant.
One can show that for not-too-small $\lambda_T$, i.e.\ 
as long as $\lambda_T^2 > 2\,\lambda_+ \lambda_-$, it is 
in fact always smaller than the singlet F-terms,
$\left|F_T\right| < \left|F_{Z_\pm}\right|$.}


Setting $M$ to its VEV, $\left<M\right> = 0$, and using that
$ \left<M_+\right> \left<M_-\right> = \Lambda^2$ for $\lambda_T \rightarrow\infty$,
we now have
\begin{align}
W_{\rm eff} \simeq \lambda_+ \Lambda \, M_+ \, Z_- + 
\lambda_- \Lambda \, M_- \, Z_+  \,, 
\label{eq:WMpm}
\end{align}
where the charged meson fields $M_+$ and $M_-$ are to be expanded around their
VEVs as follows,
\begin{align}
M_+ = \left<M_+\right>  e^S \,, \quad
M_- = \left<M_-\right>  e^{-S} = \frac{\Lambda^2}{\left<M_+\right>} e^{-S} = \frac{\Lambda^2}{M_+} \,.
\label{eq:MVEVpara}
\end{align}
This illustrates once more that the deformed moduli constraint is also satisfied on
the level of the chiral superfields, $M_+ M_- = \Lambda^2$, and not only on the level
of the scalar VEVs.
In order to actually calculate $\left<M_+\right>$ and $\left<M_-\right>$,
we need to minimize the F-term scalar potential resulting from Eq.~\eqref{eq:WMpm}
in combination with the D-term scalar potential in Eq.~\eqref{eq:VD}, while taking
into account that $M_- = \Lambda^2/M_+$ for all values of $M_+$.
Let us assume for a moment that $\left<Z_\pm\right> = 0$ (we will justify
this assumption further below in Sec.~\ref{subsec:spectrum}). 
We then find for the VEVs of the charged meson fields
\begin{align}
\big< \left|M_\pm\right|^2 \big> =
\frac{\lambda_\mp}{\lambda_\pm}  \Lambda^2 \left[1 + \frac{g^2}{2}
\left(\frac{\lambda_\pm}{\lambda_\mp^3} - \frac{\lambda_\mp}{\lambda_\pm^3}\right) +
\mathcal{O}\left(g^4\right)\right]\,.
\label{eq_mvevs}
\end{align}
As we are only interested in the limit of a weakly gauged $U(1)$ symmetry, $g \ll 1$,
we can safely neglect all higher-order corrections in $g$.
In the generic case, we expect the Yukawa couplings $\lambda_+$ and $\lambda_-$
to (at least slightly) differ from each other, $\lambda_+ \neq \lambda_-$,
so that the VEVs of the two mesons $M_+$ and $M_-$ do not (exactly) coincide,
$\left<M_+\right>\neq \left<M_-\right>$.
According to Eq.~\eqref{eq:xiMM}, this then induces the following effective
FI parameter,
\begin{align}
\xi = 
\Lambda^2 \left( \frac{\lambda_+}{\lambda_-} - \frac{\lambda_-}{\lambda_+} \right)
\left[1 - \frac{g^2}{2}\frac{\left(\lambda_+^2 + \lambda_-^2\right)^2}
{\lambda_+^3\lambda_-^3} + \mathcal{O}\left(g^4\right)\right] \,.
\label{eq:xi}
\end{align}


This expression is the main result of our paper.
For given values of the Yukawa coupling constants $\lambda_+$ and $\lambda_-$,
it is largest in the limit $g\rightarrow0$.
This directly reflects the influence of the D-term scalar potential
in Eq.~\eqref{eq:VD} on the VEVs of the charged mesons.
The D-term scalar potential drives these VEVs towards a common value,
which results in a smaller $\xi$ parameter, as soon as the D-term
potential gains in importance.
Moreover, we find that, for  $\lambda_+$ and $\lambda_-$
of $\mathcal{O}(1)$, the magnitude of the FI parameter
is directly controlled by the dynamical scale,
$\sqrt{\left|\xi\right|}\sim\Lambda$, as anticipated.
At the same time, the VEV of the auxiliary $D$ field is suppressed by
the small $U(1)$ gauge coupling constant, $\left<D\right> = g\,\xi \sim g\,\Lambda^2$.
This needs to be compared with the magnitude of the total
F-term resulting from the effective superpotential in Eq.~\eqref{eq:WMpm},
\begin{align}
\left<\left|F\right|\right> \equiv \sqrt{\big<\left|F_{Z_+}\right|^2\big>
+ \big<\left|F_{Z_+}\right|^2\big>} = \exp\left[\frac{K_0}{2M_{\rm Pl}^2}\right] \mu^2
\,, \quad K_0 = \left<K\right> \,, \quad \mu \equiv \sqrt[4]{2\,\lambda_+\lambda_-}\,\Lambda \,,
\label{eq:IYITF}
\end{align}
which is typically of the order of the dynamical scale,
$\left<\left|F\right|\right> \sim \Lambda^2$.
In the parameter region of interest, i.e.\ for a weakly gauged $U(1)$ symmetry
and a not-too-strong hierarchy among the Yukawa coupling constants $\lambda_+$ and $\lambda_-$,
the effective D-term is hence always smaller than the IYIT F-term,
$\left<D\right> < \left<\left|F\right|\right>$, as expected, cf.\ also Eq.~\eqref{eq:DFbound}.
Note, however, that in general the IYIT model or extensions thereof
may potentially also be able to accommodate much larger
D-terms, $\left<D\right> \gtrsim \left<\left|F\right|\right>$, in case
the $U(1)$ gauge coupling is taken to a larger value,
$g \sim \lambda_\pm$.
But, as our above derivation of the $\xi$ parameter in Eq.~\eqref{eq:xi} is only
self-consistent under the assumption of a small gauge coupling constant,
$g \ll \lambda_\pm$, we cannot make any further statement as to whether
this is indeed the case or not.
Instead, we leave a study of the IYIT model (or extensions thereof) in combination
with a strongly gauged $U(1)$ flavor symmetry to future work and keep on focusing on
the weakly gauged scenario in the following.
Just as in our analysis so far, we can then continue to treat
the $U(1)$ gauge interactions as a small perturbation to the Yukawa
interactions encoded in the tree-level superpotential.


In summary, we conclude that the IYIT F-term $\left<\left|F\right|\right>$,
the FI parameter $\xi$ as well as the effective D-term
$\left<D\right>$ are all related to the dynamical scale $\Lambda$ 
of the strong $SP(1)$ gauge interactions,
\begin{align}
\left<D\right> / g \equiv \xi \sim \left<\left|F\right|\right> \sim \Lambda^2 \,.
\label{eq:hierarchy}
\end{align}
For completeness, we also mention that the dynamical scale $\Lambda$ derives in turn,
via the effect of dimensional transmutation, from the RGE running of the strong
gauge coupling constant $g_s$,
\begin{align}
\Lambda = M_{\rm Pl} \, \exp\left[-\frac{8\pi^2}{
b \, g_s\negthinspace\left(M_{\rm Pl}\right)}\right] \,, \quad
b = 3 \left(N_c + 1\right) - N_f \,, \quad N_f = N_c + 1 \,, 
\end{align}
with $b$ denoting the beta-function coefficient for the $SP(N_c)$
theory with $N_f$ flavors and where we have assumed that the
number of flavors does not change between the dynamical scale
and the Planck scale.
For $N_c = 1$, a strong gauge coupling constant of $g_s \simeq 2.0$
at the Planck scale then implies, for instance, a dynamical scale
coinciding with the scale of grand unification,
\begin{align}
N_c = 1 \,, \quad g_s\negthinspace\left(M_{\rm Pl}\right) \simeq 2.0 \,, \quad
\Lambda \simeq \Lambda_{\rm GUT}
\simeq 2.0 \times 10^{16} \,\text{GeV} \,.
\end{align}
Likewise, varying the gauge coupling constant $g_s$ between, say, $1$ and $4\pi$, we are able
to generate values of the dynamical scale $\Lambda$ ranging over almost nine order of magnitude,
\begin{align}
N_c = 1 \,, \quad g_s\negthinspace\left(M_{\rm Pl}\right) \simeq 1 \, .. \, 4\pi \,, \quad
\Lambda \simeq 6.5 \times 10^9 \,\text{GeV} \,..\, 2.1 \times 10^{18} \,\text{GeV} \,.
\label{eq:Lambdavalues}
\end{align}
Together, Eqs.~\eqref{eq:hierarchy} and \eqref{eq:Lambdavalues} thus illustrate
explicitly that our dynamical mechanism for the generation
of an effective FI-term is capable of yielding $\xi$ values in a much larger
range than the conventional string construction based on the Green-Schwarz
mechanism.
This is a major advantage of our dynamical, field theory-based scenario.


\subsection{Consistent embedding into supergravity}
\label{subsec:embedding}


In the above derivation of the effective FI parameter $\xi$, we
solely worked in the limit of global SUSY and completely
neglected all SUGRA effects.
This immediately gives rise to two questions:
(i) What are the quantitative changes in our result for the parameter $\xi$
once we include higher-dimensional SUGRA corrections?
And more importantly, (ii) are we at all allowed to couple our globally
supersymmetric $SP(1)$ model to SUGRA without running into such conceptional
problems as we discussed them in the introduction?
Assuming the answer to the second question is positive,
the first question is easy to answer.
Simply based on dimensional analysis, we expect all SUGRA corrections
to our above expressions to be suppressed by at least one power of
the ratio $\Lambda/M_{\rm Pl}$.
For not-too-large values of the dynamical scale, $\Lambda \lesssim 10^{-1}\,M_{\rm Pl}$,
all SUGRA corrections are therefore well under control.
In order to answer the second question, it is sufficient to note
that the superpotential as well as the K\"ahler potential of our
original high-energy $SP(1)$ theory are $U(1)$ gauge-invariant
by construction.
The FZ-multiplet is therefore always well-defined along the entire RGE flow
and we do not have to worry about any complications when coupling our model
to SUGRA.
However, in order to make the virtues of our dynamical mechanism more explicit,
it turns out to be useful to examine the K\"ahler potential
of the low-energy effective theory in a bit more detail.
More precisely, we shall now identify and discuss the analog of the K\"ahler
potential $K_{\rm FI}$ in Eq.~\eqref{eq:LagFI} in our model.


The canonical K\"ahler potential for the meson fields $M_\pm$
in global SUSY is given as follows,
\begin{align}
K = M_+^\dagger\, e^{2gV} M_+ + M_-^\dagger\, e^{-2gV} M_- \,.
\label{eq:KMM}
\end{align}
Parameterizing the fluctuations of $M_+$ and $M_-$ around the low-energy vacuum
as in Eq.~\eqref{eq:MVEVpara}, this K\"ahler potential can be written as
\begin{align}
K = K_0 + 2g \left[\big< \left|M_+\right|^2 \big> - \big< \left|M_-\right|^2 \big>\right]
\left[V + \frac{1}{2g}\left(S + S^\dagger\right)\right] + \dots \,, 
\label{eq:KVSS}
\end{align}
where $K_0 = \big< \left|M_+\right|^2 \big> + \big< \left|M_-\right|^2 \big>$
and with the ellipsis denoting higher-order terms in the linear combination
$V + \left(S + S^\dagger\right)/(2g)$.
Of course, the prefactor of the above linear term is nothing but our
effective FI parameter $\xi$, cf.\ Eq.~\eqref{eq:xiMM}.
This leads us to identify $K_{\rm FI}$ in our model with
\begin{align}
K_{\rm FI} = -2g\,\xi \left[V + \frac{1}{2g}\left(S + S^\dagger\right)\right] \,,
\label{eq:KFI}
\end{align}
which is of exactly the same form as the K\"ahler potential
in Eq.~\eqref{eq:KFIstring} resulting from the Green-Schwarz
mechanism in string theory.
First of all, this illustrates that the field $S$ indeed represents in fact
nothing else than the Goldstone multiplet associated with the spontaneous
breaking of the $U(1)$ symmetry, which renders the vector multiplet
$V$ massive via the (affine) Abelian Higgs mechanism,
$V\rightarrow V_M = V + \left(S + S^\dagger\right)/(2g)$.
But more than that, the similarity between our $K_{\rm FI}$ in Eq.~\eqref{eq:KFI}
and the stringy K\"ahler potential in Eq.~\eqref{eq:KFIstring} also shows that we
are able to consistently couple our effective FI-term to SUGRA
for the same reason as in string theory---in contrast to the K\"ahler potential
in Eq.~\eqref{eq:LagFI}, our K\"ahler potential $K_{\rm FI}$ in
Eq.~\eqref{eq:KFI} does not violate the gauge invariance of the FZ-multiplet.
As the field $S$ shifts under super-gauge transformations,
$S \rightarrow S - 2g\Psi$, it exactly compensates for
the gauge shift in the vector field, $V \rightarrow V + \Psi + \Psi^\dagger$.
This renders $K_{\rm FI}$ gauge-invariant, $K_{\rm FI} \rightarrow K_{\rm FI}$,
which ultimately also preserves the gauge invariance of the FZ-multiplet.
The advantage of our dynamically generated FI-term compared to its
stringy counterpart, though, is that the modulus field, i.e.\ the real
scalar DOF contained in $S+S^\dagger$, is always automatically
stabilized by the SUSY-breaking F-terms arising in our strongly coupled theory.
Unlike in the case of the stringy modulus field $\Phi$ in Eq.~\eqref{eq:KFIstring},
no extra mechanism is therefore required to stabilize $S$.
In particular, thanks to the spontaneous breaking of SUSY, the mass of
the field $S$ always ends up being parametrically larger than the mass of the
$U(1)$ vector boson $A$.
At energies below the SUSY breaking scale, but above the vector boson mass,
our effective FI-term hence manages to resemble a genuine, constant FI-term
to very good approximation.
Our dynamical mechanism is thus not only consistent in the context of SUGRA,
but also appears to be very promising from a phenomenological point of view.


\subsection{Mass spectrum and stabilization of the sgoldstino direction}
\label{subsec:spectrum}


To quantify the above statements about the mass of the field $S$
and its relation to the vector boson mass, let us explicitly reformulate
our $SP(1)$ SUSY breaking model in terms of the Goldstone field $S$ and
the singlet fields $Z_+$ and $Z_-$.
This will also allow us to eventually prove that $Z_+$ and $Z_-$ are
indeed stabilized around their origin, cf.\ the comment above Eq.~\eqref{eq_mvevs}.
Using Eqs.~\eqref{eq:WMpm}, \eqref{eq:MVEVpara} and \eqref{eq_mvevs}, the
effective superpotential may be rewritten as,%
\footnote{Unless stated otherwise, we will neglect all effects due
to the $U(1)$ gauge interactions in the following and simply work in zeroth
order in the gauge coupling.
Similarly as before, we will also neglect all SUGRA corrections.}
\begin{align}
W_{\rm eff} \simeq \mu^2 \left[\cosh\left(S\right)X -
\sinh\left(S\right)Y\right] \,, \quad  \mu = \sqrt[4]{2\lambda_+\lambda_-} \, \Lambda \,,
\label{eq:Wcoshsinh}
\end{align}
where we have introduced the following two linear combinations of the fields
$Z_+$ and $Z_-$, 
\begin{align}
X = \frac{1}{\sqrt{2}} \left(Z_+ + Z_-\right) \,, \quad
Y = \frac{1}{\sqrt{2}} \left(Z_+ - Z_-\right) \,.
\label{eq:XY}
\end{align}
As evident from Eq.~\eqref{eq:Wcoshsinh}, for a generic value of the Goldstone
field $S$, both $X$ and $Y$ possess nonvanishing F-terms.
The sum of the absolute values squared of these F-terms then yields
the scalar potential for the complex scalar contained in $S$,
\begin{align}
V_S = \left|F_X\right|^2 + \left|F_Y\right|^2 = \mu^4\cosh\big(\sqrt{2}\,c\big) \,,
\quad S = \frac{1}{\sqrt{2}}\left(c + i \varphi\right) \,.
\label{eq:VS}
\end{align}
We hence find that $c$, the real component of the complex scalar in $S$,
is stabilized at the origin, while $\varphi$, the imaginary component of
the complex scalar in $S$, turns out to be a massless flat direction.
This is consistent with the fact that $\varphi$ is to be identified with
the actual Goldstone phase that is absorbed by the $U(1)$ vector field $A$
upon the spontaneous breaking of the $U(1)$ symmetry.
Owing to ordinary gauge invariance, we are then always allowed to shift
$\varphi$ to zero, so that the entire $S$ multiplet vanishes, $S = 0$.
This gauge choice corresponds to unitary gauge, in which $S$ disappears
from the superpotential, because it is eaten by the vector multiplet $V$.


In unitary gauge, the superpotential then takes the following,
particularly simple form,
\begin{align}
W_{\rm eff} \simeq \mu^2 X \,.
\label{eq:WX}
\end{align}
To study the particle spectrum of our $SP(1)$ theory in this gauge,
one can no longer use the standard representation of the $U(1)$ vector
multiplet that is commonly employed in Wess-Zumino gauge.
Instead, one needs to know the actual Lagrangian of our Abelian Higgs model
in unitary gauge, which includes, \textit{inter alia}, all interactions
of the fields $X$ and $Y$ with the massive vector multiplet $V_M$ that
are encoded in the K\"ahler potential.
This Lagrangian has been derived and discussed in detail in
Ref.~\cite{Buchmuller:2012wn}.
Alternatively, we can, however, also simply keep the field $S$ in
the superpotential and perform all calculations in Wess-Zumino gauge.
The conceptional difference between these two approaches then is that
the Goldstone multiplet is regarded as a set of gauge DOFs in the former case,
while it is regarded as a set of matter DOFs in the latter case.
Of course, both approaches are guaranteed to lead to the same physical
results by virtue of the Goldstone boson equivalence theorem.
Based on the Lagrangian stated in
Ref.~\cite{Buchmuller:2012wn}, we have also
confirmed the equivalence between both approaches by means of
an explicit calculation.


For the purposes of the discussion in the present paper, we shall
now continue our analysis in Wess-Zumino gauge.
In doing so, we are free to focus, without loss of generality, on
fluctuations of the Goldstone field around unitary gauge, $S=0$.
For $S$ values close to $0$, the superpotential in Eq.~\eqref{eq:Wcoshsinh} can
then be expanded in powers of $S$ up to second order as follows
\begin{align}
W_{\rm eff} \simeq \mu^2 \left[X - S\,Y + \frac{1}{2} S^2\,X +
\mathcal{O}\left(S^3\right)\right] \,.
\label{eq:WXYS}
\end{align}
Likewise, we may expand the K\"ahler potential in Eq.~\eqref{eq:KMM}
up to second order in $S$.
Setting the vector field to its VEV, $\left<V\right> = 0$,
we find the following non-holomorphic terms in $K$, 
\begin{align}
K \supset K_0 \left[1 + \left|S\right|^2 + \mathcal{O}\big(\left|S\right|^4\big) \right] \,,
\quad K_0 = \left(\frac{\lambda_-}{\lambda_+} + \frac{\lambda_+}{\lambda_-}\right) \Lambda^2 \,.
\end{align}
The field $S$ thus does not possess a canonical kinetic term.
In order to canonically normalize it, we have to perform a field redefinition,
$S \rightarrow K_0^{-1/2}S$, such that
\begin{align}
K \rightarrow & \: K_0 + \left|S\right|^2 + \mathcal{O}\big(\left|S\right|^4\big) \,,
\label{eq:WOR}\\ \nonumber
W_{\rm eff} \rightarrow & \: \mu^2 X - m \, S \, Y + \frac{m^2}{2\mu^2}\,S^2 \,X +
\mathcal{O}\left(S^3\right)  \,,
\end{align}
with the mass parameter $m$ being defined as,
\begin{align}
m = \frac{\mu^2}{K_0^{1/2}} =
\lambda_h \, \Lambda \,, \quad \lambda_h = \left[\frac{1}{2}\left(
\frac{1}{\lambda_+^2} + \frac{1}{\lambda_-^2}\right)\right]^{-1/2} \,,
\label{eq:mpara}
\end{align}
where $\lambda_h$ denotes the positive square root of the harmonic mean
of $\lambda_+^2$ and $\lambda_-^2$.
Again, we can immediately infer several important results
simply from the form of the superpotential.


(i) The O'Raifeartaigh-like structure of the superpotential
is clearly evident.
The field $X$ eventually turns out to be the only field with a nonvanishing $F$-term.%
\footnote{According to Eq.~\eqref{eq:WXYS}, $\left<S^2\right> = -2$
appears, in fact, to represent a viable vacuum configuration as well.
In this vacuum, $F_X$ would then vanish and SUSY would instead be broken
by the F-term belonging to the field $Y$.
However, this is a fallacy, since Eq.~\eqref{eq:WXYS} only holds in the limit
of a small VEV for the Goldstone field, $S \ll 1$.}
Hence, in unitary gauge, it directly corresponds to the goldstino multiplet associated
with the spontaneous breaking of SUSY, cf.\ also Eq.~\eqref{eq:WX}.
The complex scalar contained in $X$, the sgoldstino field,
is typically very light at tree level, which is why it represents
what may be referred to as a pseudomodulus.
The only contributions to its tree-level mass originate from higher-dimensional
operators in the K\"ahler potential and are thus expected to be very small.
At the same time, the fermionic component of $X$, the initially massless
goldstino field $\tilde{x}$, is eaten by the gravitino upon the spontaneous
breaking of SUSY, representing its longitudinal DOFs thereafter.
Hence, the mass of the goldstino $\tilde{x}$ eventually ends up
corresponding to the gravitino mass $m_{3/2}$,
\begin{align}
m_{\tilde{x}} \equiv m_{3/2} =
\exp\left[\frac{K_0}{2M_{\rm Pl}^2}\right] \frac{W_0}{M_{\rm Pl}^2} \,, \quad W_0 = \left<W\right> \,.
\end{align}
Here, $W_0$ denotes a constant in the superpotential that is generated in the
course of spontaneous $R$ symmetry breaking at high energies.
It is tuned against the total SUSY breaking scale $\Lambda_{\rm SUSY}$ in the scalar
potential, so as to realize an almost vanishing cosmological constant in the true vacuum.%
\footnote{In Ref.~\cite{Schmitz:2014xx}, we will present an extension of our general
$SP(N_c)$ model, based on the idea of conformal SUSY breaking~\cite{Ibe:2008td}, that
incorporates a dynamical explanation for this fine-tuning at least at the classical level.}
This tuning implies the following phenomenological relation between $m_{3/2}$
and $\Lambda_{\rm SUSY}$,
\begin{align}
m_{3/2} = \frac{\Lambda_{\rm SUSY}^2}{\sqrt{3} M_{\rm Pl}} \,, \quad 
\Lambda_{\rm SUSY}^2 = \sqrt{\big<\left|F_{\rm tot}\right|^2\big> + \frac{1}{2}\left<D^2\right>} \,.
\label{eq:m32LambdaSUSY}
\end{align}
Note that the above $\left|F_{\rm tot}\right|$ does not necessarily need to
coincide with the IYIT F-term $\left|F\right|$ in Eq.~\eqref{eq:IYITF},
since additional hidden sectors might still provide further sources of
F-term-driven SUSY breaking.
For our purposes, we can therefore treat the gravitino mass as a free parameter,
which may or may not be determined by the dynamical scale in the strongly
coupled sector.


For now, we would like to stress only two points:
First, if there should be no further sources of SUSY breaking except for those coming
from our strongly coupled sector, the SUSY breaking scale $\Lambda_{\rm SUSY}$ is
also determined by the dynamical scale $\Lambda$.
On the other hand, in presence of additional sources of SUSY breaking,
$\Lambda$ represents a lower bound for $\Lambda_{\rm SUSY}$,
cf.\ also Eq.~\eqref{eq:hierarchy},
\begin{align}
\Lambda_{\rm SUSY} \gtrsim \left<\left|F\right|\right> \sim \xi \sim
\Lambda \,, \quad m_{3/2} \gtrsim \frac{\Lambda^2}{\sqrt{3} M_{\rm Pl}} \,.
\end{align}
Therefore, if we really aim at generating a $\xi$ value of
the order of $\Lambda_{\rm GUT}$, we automatically let ourselves
in for an extremely large SUSY breaking scale.
In the case of such a large scale $\Lambda_{\rm SUSY}$, all 
effects of SUSY decouple from low-energy physics and we could
have no hope to solve any problems of the standard model
by means of SUSY---this may or may not be regarded as a drawback.
Alternatively, we may simply envision the dynamical generation
of an effective $\xi$ parameter of a much smaller magnitude.
Then, also the SUSY breaking scale might end up lying
in a phenomenologically more attractive range.
Our second comment pertains the gravitino mass.
Without any extra sources of SUSY breaking present,
we know that $m_{3/2}$ takes a value of
$\mathcal{O}\left(\Lambda^2 / M_{\rm Pl}\right)$,
cf.\ Eq.~\eqref{eq:m32LambdaSUSY}.
All SUGRA corrections proportional to $m_{3/2}$ in our above analysis are
then be suppressed by the ratio $\Lambda / M_{\rm Pl}$.
This is exactly what we already estimated
simply based on dimensional analysis at the beginning of Sec.~\ref{subsec:embedding}.
For further comments on the mediation of SUSY breaking to the visible
sector in our dynamical model, cf.\ Sec.~\ref{subsec:susybreaking}.


(ii) The singlet field $Y$ shares a large supersymmetric Dirac mass term with the Goldstone
field $S$.
Both fields are hence stabilized around the origin, $\left<Y\right> = \left<S\right> = 0$.
Here, the fact that the linear combination $Y = \left(Z_+ - Z-\right)/\sqrt{2}$ vanishes
incidentally implies that the singlet fields $Z_+$ and $Z_-$ are bound to take the same
value in the true vacuum, $\left<Z_+\right> = \left<Z_-\right>$, independently of the
actual VEV of the linear combination $X = \left(Z_+ + Z-\right)/\sqrt{2}$.
Interestingly enough, this means that the contributions from the VEVs
of the fields $Z_+$ and $Z_-$ to the expectation value of the auxiliary
$D$ field, $\left<D\right> \supset g\,\big(\big<\left|Z_-\right|^2\big> -
\big<\left|Z_+\right|^2\big>\big)$, must necessarily cancel.

Upon closer inspection, we find that the two chiral fields $S$ and $Y$
give in fact rise to four real scalar as well as to two fermionic mass eigenstates,%
\footnote{The mixing between the DOFs contained in $S$ and $Y$
vanishes for $X = 0$.
At $X = 0$, we then have $S\sim(c,\varphi,\tilde{s})$ and $Y\sim(y_-,y_+,\tilde{y})$.
In the rigid limit, $X$ vanishes in the true vacuum (as we will see shortly)
and, hence, the role of the Goldstone multiplet is exclusively played by the field $S$. 
In SUGRA, however, $\left<X\right>$ turns out to be of
$\mathcal{O}\left(m_{3/2}\right)$, so that the Goldstone DOFs
end up corresponding to linear combinations of the DOFs in $S$ and $Y$.}
\begin{align}
\left\{S,Y\right\}_{\text{bosonic DOFs}} \rightarrow \left\{c,\varphi,y_-,y_+\right\}
\,, \quad
\left\{S,Y\right\}_{\text{fermionic DOFs}} \rightarrow \left\{\tilde{s},\tilde{y}\right\} \,.
\end{align}
Almost all of these fields possess an effective mass that depends on the
complex scalar contained in the field $X$.
Around the origin, i.e.\ at $\left<S\right> = \left<Y\right> = 0$, we find for
the effective masses squared
\begin{align}
m_c^2 = & \: m^2 \left[\frac{3}{2} + \frac{m^2}{2\,\mu^4} \left|X\right|^2
+ \left(\frac{1}{4} + \frac{3\,m^2}{2\,\mu^4}\left|X\right|^2 + \frac{m^4}{4\,\mu^8}
\left|X\right|^4 \right)^{1/2}\right]
\,, \quad m_\varphi^2 =  0 \,, 
\label{eq:masses2}\\ \nonumber
m_{y_-}^2 = & \: m^2 \left[\frac{3}{2} + \frac{m^2}{2\,\mu^4} \left|X\right|^2
- \left(\frac{1}{4} + \frac{3\,m^2}{2\,\mu^4}\left|X\right|^2 + \frac{m^4}{4\,\mu^8}
\left|X\right|^4 \right)^{1/2}\right]
\,, \quad m_{y_+}^2 =  m^2 + \frac{m^4}{\mu^4} \left|X\right|^2  \,,
\\ \nonumber
m_{\tilde{s}}^2 = & \: m^2 \left[1 +\frac{m^2}{2\,\mu^4} \left|X\right|^2
+ \left(\frac{m^2}{\mu^4} \left|X\right|^2 + \frac{m^4}{4\,\mu^8} \left|X\right|^4\right)^{1/2}\right] \,, 
\\ \nonumber
m_{\tilde{y}}^2 = & \: m^2 \left[1 + \frac{m^2}{2\,\mu^4} \left|X\right|^2
- \left(\frac{m^2}{\mu^4} \left|X\right|^2 + \frac{m^4}{4\,\mu^8} \left|X\right|^4 \right)^{1/2}\right] \,.
\end{align}
Here, $m_{\tilde{s}}^2$ and $m_{\tilde{y}}^2 $ are understood to denote the Majorana
masses for the two Majorana fermions $\tilde{s}$ and $\tilde{y}$ at $X \neq 0$.
On the other hand, at $X = 0$, the fermions $\tilde{s}$ and $\tilde{y}$
share a common mass $m$, so that they can
be regarded as forming a Dirac fermion together.
More generally speaking, evaluated at $X = 0$, the effective masses
in Eq.~\eqref{eq:masses2} reduce to the following vacuum masses,
\begin{align}
X = 0 \,: \quad\quad
m_c^2 = 2\, m^2 \,, \quad m_\varphi^2 = 0 \,, \quad
m_{y_-}^2 = m_{y_+}^2 = m^2 \,, \quad m_{\tilde{s}}^2 = m_{\tilde{y}}^2 = m^2 \,.
\label{eq:X0masses}
\end{align}


As already mentioned below Eq.~\eqref{eq:VS}, the Goldstone phase $\varphi$
turns out to be exactly massless and the real scalar $c$ receives a large mass
and is thus stabilized at the origin, $\left<c\right> = 0$.
Here, the scalar $c$ corresponds to the real scalar in the massive vector
multiplet $V_M \sim \left(c,\lambda,\tilde{s},A\right)$ in unitary gauge.
In addition, it plays the role of the real scalar modulus field in
our construction of the effective FI-term in Eq.~\eqref{eq:KFI}.
As anticipated in Sec.~\ref{subsec:embedding}, we now see explicitly 
that is indeed always automatically stabilized due to the
F-term-driven SUSY breaking inherent in our superpotential.
In particular, for a weakly gauged $U(1)$ symmetry, its F-term-induced
mass always easily satisfies the constraint formulated in the
introduction, cf.\ Sec.~\ref{subsec:introeffectiveFI},
\begin{align}
m_c \sim \lambda_h\, \Lambda \gg  g\sqrt{\left|\xi\right|} \sim g\,\Lambda\,,
\end{align}
which guarantees that the effective FI parameter $\xi$ is not
inadvertently absorbed by the VEV of the field $c$.
Unlike in the case of string theory, we therefore do not have to
invoke any separate mechanism to stabilize the modulus field.
The dynamical stabilization of the modulus is already built into
our model.
Furthermore, the modulus mass is not directly tied to the gravitino
mass, which leaves more freedom in the construction of realistic
scenarios.
In fact, barring additional sources of SUSY breaking, $m_{3/2}$
is typically much smaller than the modulus mass,
$m_{3/2} \sim \Lambda^2 / M_{\rm Pl} \ll m_c$, cf.\ Eq.~\eqref{eq:m32LambdaSUSY}.
These results regarding the scalar field $c$ represent definite advantages
of our dynamically generated FI-term over the stringy construction
based on the Green-Schwarz mechanism.


The masses of the scalar $c$ as well as of the Majorana fermion
$\tilde{s}$ now need to be compared with the masses of the vector
boson $A$ as well as of the gaugino $\lambda$.
To compute these masses, we simply have to inspect the K\"ahler potential
in Eq.~\eqref{eq:KVSS}, which can also be written as
\begin{align}
K = K_0 - 2g\, \xi \, V_M + m_V^2 V_M^2 + \mathcal{O}\left(V_M^3\right) \,,
\quad V_M = V + \frac{1}{\sqrt{2}\,m_V}\big(S + S^\dagger\big) \,, \quad
m_V^2 = 2g^2 K_0 \,.
\label{eq:KVM}
\end{align}
As expected, we thus find that the vector boson $A$ acquires a mass
$m_A \equiv m_V \sim g\,\Lambda$.
With respect to the gaugino $\lambda$, things are a little bit more complicated.
Next to the mass for the vector boson $A$, the K\"ahler potential in Eq.~\eqref{eq:KVM} 
also induces a Dirac mass for the two fermions $\lambda$ and $\tilde{s}$.
Together with the Dirac mass $m$ for $\tilde{s}$ and $\tilde{y}$,
cf.\ Eq.~\eqref{eq:X0masses}, and in the rigid limit, this then results in
three mass eigenstates with the following tree-level masses in the
true vacuum,
\begin{align}
X = 0 \,: \quad\quad
m_\lambda^2 = 0 \,, \quad m_{\tilde{s}}^2 = m_{\tilde{y}}^2 = m^2 + m_V^2 \simeq m^2 \,.
\end{align}
At $X = 0$, the fermions $\tilde{s}$ and $\tilde{y}$ therefore
form a Dirac fermion with mass $m$, while the gaugino $\lambda$ turns
out to be a Majorana fermion with vanishing tree-level mass.
Here, the masslessness of the gaugino is a direct consequence
of fermion number conservation: at $X = 0$, the Lagrangian only
contains Dirac mass terms for the three Weyl fermions $\tilde{y}$,
$\tilde{s}$ and $\lambda$ and no fermion number-violating Majorana
mass terms.
One of these three fermions therefore needs to end up being a massless
Majorana fermion, which, in our case, is the gaugino $\lambda$.
The origin of the fermion number conservation is, in turn, the continuous
global $R$ symmetry of the IYIT superpotential in the rigid limit,
cf.\ Eq.~\eqref{eq:Weff}, which forbids any Majorana mass terms in the
fermionic Lagrangian.


This picture of the fermionic mass eigenstates, however, receives
corrections from gravitational interactions.
In the full, locally supersymmetric case, the value of the field $X$ in the true
vacuum does not vanish, but is rather of $\mathcal{O}\left(m_{3/2}\right)$.
This induces a mass splitting $\Delta m^2$ between the fermions $\tilde{s}$ and $\tilde{y}$
of $\mathcal{O}\left(\pm m\,m_{3/2}\right)$, cf.\ Eq.~\eqref{eq:masses2}, such 
that they end up representing independent Majorana fermions after all.
At the same time, the gaugino $\lambda$ and the fermion $\tilde{y}$
acquire a common Dirac mass
$m_{\lambda\tilde{y}}\sim g \left<X\right>\sim g\,m_{3/2}$ in SUGRA
via the supersymmetric gaugino-fermion-scalar $U(1)$ gauge interactions.
After diagonalizing the fermion mass matrix, this eventually gives rise to
a Majorana mass $m_\lambda \sim m_V/m \,m_{\lambda \tilde{y}} \sim
g^2/\lambda_h \,m_{3/2}$ for the gaugino mass eigenstate.
Thus, the gaugino is, in fact, not exactly massless.
In summary, we conclude that the spontaneous breaking of SUSY in our IYIT superpotential
turns out to be responsible for a large mass splitting within the massive vector multiplet
$V_M \sim \left(c,\lambda,\tilde{s},A\right)$, 
\begin{align}
m_\lambda \sim \frac{g^2}{\lambda_h} \, m_{3/2} \,, \quad
m_A \sim g\, \Lambda \,, \quad
m_{\tilde{s}} \sim m_c \sim \lambda_h\, \Lambda \,, \quad\quad
m_\lambda \ll m_A \ll m_{\tilde{s}} \sim m_c \,.
\end{align}
In view of this result, two comments are in order.
First, for a small gauge coupling constant $g$, the $U(1)$ gaugino $\lambda$
turns out to be the lightest particle in the spectrum of our $SP(1)$ theory.
In concrete phenomenological applications of our scenario, it might therefore
be worthwhile to have a careful look at the role played by the gaugino.
Under certain circumstances, it may perhaps play the role of dark matter;
in other cases, its low mass may be used to place constraints on the dynamical
scale $\Lambda$ and/or the SUSY breaking scale $\Lambda_{\rm SUSY}$.
Second, at intermediate energies $E$ in between the two mass scales $m_V$ and $m$,
the Goldstone multiplet $S$ can be integrated out, while the gaugino
and the vector boson are still light.
At such energies, our effective FI-term then resembles a genuine,
constant FI-term to very good approximation,
\begin{align}
m_V \ll E \ll m \,: \quad \quad K_{\rm FI} \approx -2g\,\xi\,V \,.
\label{eq:fakeFI}
\end{align}
After having succeeded in generating an effective FI parameter $\xi$
by means of strong dynamics in Sec.~\ref{subsec:effectiveFI}, cf.\ Eq.~\eqref{eq:xi},
this observation is now the second main result of our paper.
Remarkably enough, our dynamical FI-Term indeed bears the potential to
imitate a constant FI-term.


(iii) We have already seen that the field $Y$ is stabilized at zero.
Hence, we have halfway proven that the VEVs of the fields $Z_+$ and $Z_-$ indeed
vanish.
What remains to be done, though, is to show that, in the rigid limit, also the
complex scalar contained in $X$, i.e.\ the sgoldstino, is stabilized at the origin.
To do so, we first point out that the effective superpotential in
Eq.~\eqref{eq:WOR} now also features a Yukawa interaction between the
goldstino field $X$ and the Goldstone field $S$ that is quadratic in $S$.
This Yukawa interaction induces an effective potential at one-loop
that lifts the pseudoflat sgoldstino direction---the Goldstone multiplet
associated with the spontaneous breaking of the $U(1)$ symmetry
therefore stabilizes the sgoldstino direction associated with the spontaneous
breaking of SUSY at the loop level.
In order to explicitly calculate the sgoldstino mass $m_X$,
we need to evaluate and differentiate
the one-loop Coleman-Weinberg potential~\cite{Intriligator:2007cp},
\begin{align}
V_{\text{CW}} = \frac{1}{64\pi^2} \,\text{STr} \,{\cal M}^4 \ln \left[ \frac{{\cal M}^2}{Q^2} \right]
\,, \quad m_X^2 = \left.\frac{\partial^2\, V_{\rm CW}}{\partial X \partial X^*}\right|_{X=0} \,.
\label{eq_CW}
\end{align}
Here, $Q$ denotes an appropriate renormalization scale and $\mathcal{M}^2$ is the direct
sum of the scalar and fermionic mass matrices squared.
It contains in particular the six sgoldstino-dependent scalar and fermion
masses in Eq.~\eqref{eq:masses2}.
Evaluating the Coleman-Weinberg potential by brute force, we obtain
the following positive sgoldstino mass squared at the origin,%
\footnote{The first derivative of $V_{\rm CW}$ w.r.t.\ $X$ vanishes.
$X = 0$ is therefore indeed a local extremum of $V_{\rm CW}$.\smallskip}
\begin{align}
m_X^2 = \frac{2\ln2-1}{16\pi^2}\left(\frac{m}{\mu}\right)^4 m^2
= \frac{2\ln2-1}{32\pi^2} \left(\frac{\lambda_h}{\lambda_g}\right)^2 \lambda_h^4 \,\Lambda^2
= \frac{2\ln2-1}{4\pi^2} \frac{\lambda_+^5\lambda_-^5}{\left(\lambda_+^2 + \lambda_-^2\right)^3}
\,\Lambda^2 \,,
\label{eq:mX2}
\end{align}
where we have introduced $\lambda_g$ to denote the positive square root of
the geometric mean of $\lambda_+^2$ and $\lambda_-^2$, cf.\ also Eq.~\eqref{eq:mpara}.
This result is independent of the renormalization scale $Q$ and, more importantly,
it is consistent with the expression found by the authors of Ref.~\cite{Chacko:1998si}
in the flavor-symmetric limit, in which all Yukawa couplings $\lambda_{ij}$ are taken
to be equal.%
\footnote{A comparison in the case of different Yukawa couplings
is not feasible, as the the corresponding expression in Ref.~\cite{Chacko:1998si}
unfortunately contains a few typos.
It is, for instance, not of the correct mass dimension.}
To see this explicitly, note that in Ref.~\cite{Chacko:1998si} the parameters $m$ and $\mu$
are given as $m=\lambda\,\Lambda$ and $\mu = \lambda^{1/2}\Lambda$, with $\lambda$
being the universal Yukawa coupling.
Setting $m$ and $\mu$ in Eq.~\eqref{eq:mX2} to these values, $m_X^2$ turns into
\begin{align}
m_X^2 = \frac{2\ln2-1}{16\pi^2} \lambda^4 \Lambda^2 \,.
\end{align}
This result is smaller than the corresponding expression in Ref.~\cite{Chacko:1998si}
by a factor of $5$, since, in contrast to Ref.~\cite{Chacko:1998si}, we have initially
decided, for simplicity, to decouple the four neutral meson-singlet pairs
$\left(M_a^0,Z_a^0\right)$ from the SUSY breaking dynamics of the charged
fields $\left(M_\pm,Z_\pm\right)$.
Moreover, the fact that we eventually find $m_X^2$ to be positive
is not a coincidence.
It is rather an implication of the global $U(1)_R$ symmetry of the
superpotential in Eq.~\eqref{eq:WOR}.
This follows from a general theorem regarding the sgoldstino mass proven
in Ref.~\cite{Shih:2007av}, which states the following:
Any SUSY breaking model of the O'Raifeartaigh type in which all chiral fields either
carry $R$ charge $0$ or $2$ is bound to lead to a positive mass squared for the
pseudoflat direction. 
Given that the fields $S$, $X$ and $Y$ in Eq.~\eqref{eq:WOR} carry $U(1)_R$ charges
$0$, $2$ and $2$, respectively, this sufficient condition is evidently fulfilled in
our case.
The positive sign of $m_X^2$ in Eq.~\eqref{eq:mX2}
is, hence, nothing but a consequence of the global $U(1)_R$ symmetry and our specific
$R$ charge assignment.


In addition to the loop-induced positive mass squared in Eq.~\eqref{eq:mX2}, the sgoldstino
mass also receives a further, uncalculable contribution from higher-dimensional operators
in the K\"ahler potential that are induced by strong-coupling effects~\cite{Chacko:1998si}.
For not-too-large Yukawa couplings, $\lambda_\pm \ll 4\pi$, this contribution is, however,
subdominant.
Finally, we therefore conclude that the sgoldstino does not destabilize the vacuum.
Instead, it is safely stabilized around the origin, $\left<X\right> = 0$.
This completes our proof that both singlet fields $Z_+$ and $Z_-$ are indeed
stabilized at the origin and justifies \textit{a posteriori} our derivation
of the effective FI parameter in Sec.~\ref{subsec:effectiveFI}.
 

\section{Generalizations}
\label{sec:generalizations}


\subsection[Dynamical D-terms based on \texorpdfstring{$SP(N_{\lowercase{c}})$}{SP(Nc)} dynamics]
{Dynamical D-terms based on \texorpdfstring{\boldmath{$SP(N_{\lowercase{c}})$}}{SP(Nc)} dynamics}


While we have introduced the IYIT model of dynamical SUSY breaking for an arbitrary
number of colors, $N_c \geq 1$, in Sec.~\ref{subsec:IYIT}, we have thereafter only
focused on the minimal case of an $SP(1)$ gauge theory in the remainder of
Sec.~\ref{sec:minimal}.
The generalization of our dynamical mechanism for the generation of an effective
FI-term to larger numbers of colors is, however, straightforward.
In this section, we will therefore only briefly summarize how the main results
found in the previous chapter translate to the more general case of strongly
coupled $SP(N_c)$ dynamics.


For a larger number of colors, $N_c \geq 2$, the tree-level superpotential
of the IYIT model in Eq.~\eqref{eq:WtreeU1} exhibits not only one, but several
anomaly-free global $U(1)$ flavor symmetries.
Each of these symmetries is equally suited to be used for
the construction of a nonvanishing FI-term.
After weakly gauging a particular $U(1)$ flavor symmetry of the tree-level
superpotential, we then have to supplement the F-term potential
with a corresponding $U(1)$ D-term potential.
In the low-energy effective theory, this D-term potential
takes the following form, cf.\ Eq.~\eqref{eq:VD},%
\begin{align}
V_D = \frac{1}{2} D^2 = \frac{g^2}{2} \left[\sum_{a = 1}^n q_a
\left(\left|M_a\right|^2 - 
\left|Z_a\right|^2 \right)\right]^2 \,, \quad
\sum_{a = 1}^n q_a = \sum_{a = 1}^n q_a^3 = 0 \,,
\label{eq:VDgen}
\end{align}
where we use $a = (i,j)$ as a collective index to label the
$n = N_f\left(2N_f - 1\right)$ different pairs of meson-singlet
charge eigenstates present at low energies.
In order to calculate the expectation value of the auxiliary $D$ field
in the true vacuum, we again need to compute the VEVs of the meson
and singlet fields.
Just as in the minimal $SP(1)$ case, all singlet fields turn out to be
stabilized at zero.
This also includes the sgoldstino direction, which again receives a
sufficiently large mass at the loop level.
To facilitate our analysis of the meson VEVs, we again assume a particular
hierarchy among the Yukawa coupling constants $\lambda_{ij}$, such that
all mesons $M^{ij}$ with $\left|i-j\right|\neq 1$ vanish in the true
vacuum.
This renders $M^{ij}$ an antisymmetric tridiagonal matrix,
\begin{align}
M^{ij} = M_\alpha \, J^{ij} \,, \quad \alpha = (2i-1,2i) \,,
\quad \alpha = 1,2,..,N_f \,,
\label{eq:MijMa}
\end{align}
where the $J^{ij}$ represent the entries of the symplectic form
$J = \mathbbm{1}_{N_f}\otimes i\sigma_2$.
Here, $\mathbbm{1}_{N_f}$ stands for the $N_f$-dimensional
unit matrix and $\sigma_2$ is the second Pauli matrix.
The form $J$ is constructed such that it has unit Pfaffian, $\text{Pf}(J) = 1$.
Making use of this ansatz, the F-term potential deriving from the
superpotential in Eq.~\eqref{eq:Weff} is then minimized by the following
nonzero VEVs,
\begin{align}
\big<\left|M_\alpha\right|^2\big>^{1/2} = \frac{\lambda_g}{\lambda_\alpha} \, \Lambda
+ \mathcal{O}\left(g^2\right) \,, \quad
\lambda_g = \prod_{\alpha=1}^{N_f} \lambda_\alpha^{1/N_f} \,, \label{eq:MijVEV}
\end{align}
with $\lambda_g$ denoting the geometric mean of the Yukawa couplings
$\lambda_{12}$, $\lambda_{34}$, .., $\lambda_{2N_f-1,2N_f}$,
Also, note that now we are completely neglecting all corrections
to the meson VEVs coming from the $U(1)$ gauge interactions.
Inserting the meson VEVs in Eqs.~\eqref{eq:MijMa} and \eqref{eq:MijVEV}
into the superpotential in Eq.~\eqref{eq:Weff},
we are then able to identify the goldstino field $X$,
\begin{align}
M^{ij} \rightarrow \frac{\lambda_g}{\lambda_{ij}} \, \Lambda \, J^{ij} \,, \quad
W_{\rm eff} \rightarrow \mu^2 X \,, \quad
\mu^2 = N_f^{1/2}  \lambda_g\, \Lambda^2 \,, \quad
X = \frac{1}{N_f^{1/2}} \sum_{\alpha=1}^{N_f} Z_\alpha \,,
\end{align}
which generalizes the corresponding expressions found in the 
previous chapter, cf.\ Eqs.~\eqref{eq:Wcoshsinh}, \eqref{eq:XY} 
and \eqref{eq:WOR}.
Similarly, plugging the meson VEVs into the D-term potential in
Eq.~\eqref{eq:VDgen}, allows us to deduce the generalized expression
for the FI parameter $\xi$, cf.\ Eq.~\eqref{eq:xi},
\begin{align}
\xi = - \sum_{a = 1}^{n} q_a \left<\big|M_a\right|^2\big> =
- \lambda_g^2\,\Lambda^2 \sum_{\alpha = 1}^{N_f} \frac{q_\alpha}{\lambda_\alpha^2}
+ \mathcal{O}\left(g^2\right) \,.
\label{eq:xigen}
\end{align}
For generic Yukawa coupling constants $\lambda_\alpha$, we hence find again a nonzero
effective FI-term, the magnitude of which is determined by the dynamical scale,
$\xi \sim \Lambda^2$.
The corresponding effective D-term $\left<D\right>$ is therefore again suppressed 
compared to the IYIT F-term $\left<\left|F\right|\right>$, cf.\ Eq.~\eqref{eq:hierarchy},
\begin{align}
\left<\left|F\right|\right> \simeq \mu^2 \sim \Lambda^2 \gg
\left<D\right> \equiv g\, \xi \sim g\,\Lambda^2 \,.
\end{align}


The fluctuations of the meson fields with nonzero VEV around the low-energy
vacuum, cf.\ Eq.~\eqref{eq:MijVEV}, can be parametrized in a similar way
as in Eq.~\eqref{eq:MMMS}.
Schematically, we have
\begin{align}
M_\alpha = \mathcal{O}_{\alpha\beta}\left(S_1, S_2, .., S_{N_f-2}\right)
\frac{1}{\lambda_\beta}\left[\lambda_g\,\Lambda + M \right] e^{q_\beta S_0} \,,
\end{align}
with $M$ and $S_0$ denoting two chiral superfields of mass dimension $1$ and $0$,
respectively, and where $\mathcal{O}$ is an element of
$SO(N_f,\mathbb{C})/\left[SO(N_f -1,\mathbb{C})\times U(1)_{\mathbb{C}}\right]$,
which is uniquely determined in terms of the $N_f-2$ superfield-valued phases
$S_1$, $S_2$, .., $S_{N_f-2}$.
The field $M$ couples again to the Lagrange multiplier $T$ and hence decouples
in the limit of a large Yukawa coupling $\lambda_T$, i.e.\ once we enforce
the deformed moduli constraint to be fulfilled exactly.
Meanwhile, the field $S_0$ can again be identified as the Goldstone multiplet that renders
the $U(1)$ vector field $V$ massive upon the spontaneous breaking of the $U(1)$
symmetry.
From its interaction with the singlet fields $Z_\alpha$ in the superpotential,
it acquires a supersymmetric Dirac mass $m$, which directly generalizes the
mass parameter $m$ introduced in Sec.~\ref{subsec:spectrum},
cf.\ Eq.~\eqref{eq:mpara},
\begin{align}
m = \lambda_h\,\Lambda \,, \quad \lambda_h =
\left[\frac{1}{N_f}\sum_{i=1}^{N_f} \frac{q_\alpha^2}{\lambda_\alpha^2}\right]^{-1/2} \,.
\end{align}
Hence, the real modulus field $c \in S_0$, which appears in our
construction of the effective FI-term in the K\"ahler potential,
cf.\ Eq.~\eqref{eq:KFI}, is again parametrically heavier than
the vector boson $A$,
\begin{align}
m_A^2 \equiv m_V^2 = 2g^2 
\lambda_g^2\,\Lambda^2 \sum_{i = 1}^{N_f} \frac{q_\alpha^2}{\lambda_\alpha^2} \,.
\end{align}
At energies $E$ in between the mass scales $m$ and $m_V$, our effective
FI-term in Eq.~\eqref{eq:xigen} therefore resembles once again a
genuine, constant FI-term, cf.\ Eq.~\eqref{eq:fakeFI}.
In summary, we thus conclude that all of our main results derived in the
special case of an $SP(1)$ theory readily carry over to the general scenario
of strongly interacting $SP(N_c)$ dynamics.
This illustrates that our dynamical mechanism for the generation of
an effective FI-term, which we introduced in Sec.~\ref{sec:minimal} only by
reference to the simplest case of strongly coupled $SP(1)$ dynamics, does,
in fact, not depend on any peculiarities of the IYIT model
in its minimal realization.
Instead, it is applicable in the full IYIT model for an arbitrary
number of colors.


\subsection[Dynamical D-terms based on \texorpdfstring{$SU(N_{\lowercase{c}})$}{SU(Nc)} dynamics]
{Dynamical D-terms based on \texorpdfstring{\boldmath{$SU(N_{\lowercase{c}})$}}{SU(Nc)} dynamics}


The results of the previous section immediately entail the question whether
our dynamical mechanism could possibly also be implemented in other models
of dynamical SUSY breaking.
In this section, we shall briefly demonstrate that, properly taking into account
gravitational corrections, it actually turns out
to be impossible to dynamically generate an effective FI-term in
the context of DSB models based on $SU(N_c)$ dynamics.
This will help us formulate a general requirement pertaining the structure
of candidate DSB models that needs to be satisfied, so that there is a
chance of successfully accommodating an effective FI-term in the full SUGRA theory.
The concrete investigation of further alternative DSB models is left for future work.


The generalization of the IYIT model to $SU(N_c)$ dynamics is based on
supersymmetric QCD (SQCD) with $N_f = N_c$ flavors~\cite{Seiberg:1994pq}.
Here, we will only consider $N_c$ values larger than $2$, as the $SU(2)$
theory is equivalent to the $SP(1)$ theory discussed in Sec.~\ref{sec:minimal}.
Now, every flavor is comprised of a pair of a chiral quark and a chiral
antiquark field, $Q^i$ and $\bar{Q}^{\bar{\imath}}$, which transform in the
fundamental and antifundamental representations of $SU(N_c)$, respectively.
The DOFs of the low-energy effective theory correspond to a set of $N_f^2 +2$
gauge-invariant composite fields: $N_f^2$ mesons $M^{i\bar{\jmath}}$ as well
as a baryon $B$ and an antibaryon $\bar{B}$,
\begin{align}
M^{i\bar{\jmath}} = \frac{Q^i \bar{Q}^{\bar{\jmath}}}{\Lambda} \,, \quad
B = \epsilon_{i_1 i_2 .. i_{N_c}} \frac{Q^{i_1}Q^{i_2}..Q^{i_{N_c}}}{\Lambda^{N_c-1}} \,, \quad
\bar{B} = \epsilon_{\bar{\imath}_1 \bar{\imath}_2 .. \bar{\imath}_{N_c}}
\frac{\bar{Q}^{\bar{\imath}_1}\bar{Q}^{\bar{\imath}_2}..
\bar{Q}^{\bar{\imath}_{N_c}}}{\Lambda^{N_c-1}} \,, \quad
\end{align}
which are again subject to a quantum mechanically deformed moduli
constraint~\cite{Seiberg:1994bz},
\begin{align}
B\bar{B} + \frac{\text{det}\left(M^{i\bar{\jmath}}\right)}{\Lambda^{N_c-2}} 
= \Lambda^2 \,.
\end{align}


In this theory, we can again spontaneously break SUSY via the O'Raifeartaigh
mechanism by coupling the $2N_f$ fundamental high-energy DOFs, i.e.\ the quark and
antiquark fields $Q^i$ and $\bar{Q}^{\bar{\imath}}$, to $N_f^2 + 2$ singlet
fields, $Z_{i\bar{\jmath}}$, $Z_0$ and $\bar{Z}_0$, in the tree-level superpotential.
At low energies, this then results in the following effective superpotential
of the O'Raifeartaigh type,
\begin{align}
W_{\rm eff} \simeq & \: \lambda_T \,T 
\left[B\bar{B} + \frac{\text{det}\left(M^{i\bar{\jmath}}\right)}{\Lambda^{N_c-2}}
- \Lambda^2\right]
\label{eq:WeffSU}\\ \nonumber
+ & \: \lambda_{i\bar{\jmath}}\, \Lambda \, Z_{i\bar{\jmath}} \, M^{i\bar{\jmath}}
+ \kappa \left(\frac{\Lambda}{M_{\rm Pl}}\right)^{N_c-2} \Lambda
\, Z_0 \, B + \bar{\kappa} \left(\frac{\Lambda}{M_{\rm Pl}}\right)^{N_c-2} \Lambda
\, \bar{Z}_0 \, \bar{B} \,,
\end{align}
where $\lambda_{i\bar{\jmath}}$, $\kappa$ and $\bar{\kappa}$ denote dimensionless,
nonzero Yukawa couplings of $\mathcal{O}(1)$, which we take to be real for simplicity.
Likewise, $\lambda_T$ denotes again the normalization of the Lagrange constraint term, by means
of which we implement the deformed moduli constraint into the superpotential.
Just as for our $SP(N_c)$ theories, we assume $\lambda_T$ to be the largest coupling
in the problem.
In the low-energy vacuum of the F-term potential, all meson fields vanish
and the deformed moduli constraint ends up being satisfied due to nonzero VEVs
for the baryon and the antibaryon field,
\begin{align}
M^{i\bar{\jmath}} = 0 \,, \quad
\big<\left|B\right|^2\big>^{1/2} = \frac{\kappa_g}{\kappa} \, \Lambda \,, \quad
\big<\left|\bar{B}\right|^2\big>^{1/2} = \frac{\kappa_g}{\bar{\kappa}} \, \Lambda
\,, \quad \kappa_g = \sqrt{\kappa\bar{\kappa}} \,.
\label{eq:MBBVEVs}
\end{align}
The SUSY breaking dynamics of the general $SU(N_c)$ theory are hence very similar 
to the corresponding dynamics of the $SP(1)$ model discussed in Sec.~\ref{sec:minimal}.
Here, $B$ and $\bar{B}$ simply play the role of $M_+$ and $M_-$, while $Z_0$
and $\bar{Z}_0$ correspond to $Z_+$ and $Z_-$.
If we now weakly gauge a $U(1)$ flavor symmetry of the effective superpotential,
the VEVs of the composite states in the low-energy effective theory in
Eq.~\eqref{eq:MBBVEVs} result in an effective FI parameter $\xi$ 
of the following form,
\begin{align}
\xi = - q_B \left[\big<\left|B\right|^2\big> - 
\big<\left|\bar{B}\right|^2\big>\right] = q_B \, \kappa_g^2\,\Lambda^2
\left(\frac{1}{\bar{\kappa}^2} - \frac{1}{\kappa^2}\right) \,,
\label{eq:xiSU}
\end{align}
where $q_B$ is the charge of the baryon $B$ under the weakly gauged $U(1)$
symmetry in question.
Likewise, plugging the VEVs in Eq.~\eqref{eq:MBBVEVs} into Eq.~\eqref{eq:WeffSU},
we find the effective superpotential of our $SU(N_c)$ theories in unitary gauge,
which turns out to be very similar to the corresponding effective superpotential
that we found in the $SP(1)$ case, cf.\ Eq.~\eqref{eq:WX},
\begin{align}
W_{\rm eff} \simeq \mu^2 \, X \,, \quad
X = \frac{1}{\sqrt{2}} \left(Z_0 + \bar{Z}_0 \right) \,, \quad
\mu^2 = \sqrt{2\,\kappa\bar{\kappa}} \,
\left(\frac{\Lambda}{M_{\rm Pl}}\right)^{N_c-2} \Lambda^2 \,.
\label{eq:WXSU}
\end{align}
The crucial difference w.r.t.\ the $SP(1)$ model is that, now,
$\mu^2 \simeq \Lambda_{\rm SUSY}^2$ is suppressed
by at least one power of the ratio $\Lambda / M_{\rm Pl}$ compared
to the dynamical scale squared.
As we shall demonstrate in the following, this renders the effective FI
parameter $\xi$ in Eq.~\eqref{eq:xiSU} inconsistent.


The main obstacle to the successful generation of an effective FI-term
turns out to be the stabilization of the scalar component of the goldstino
field, i.e.\ of the pseudomodulus $X$, taking into account gravitational
corrections.
To leading order in the inverse Planck mass, $1/M_{\rm Pl}$, the one-loop
corrected potential for the sgoldstino field $X$ in SUGRA is given as
\begin{align}
V_X^{\rm eff} = \left(m_X^2 - 2\, m_{3/2}^2\right)\left|X\right|^2
- 2 \, m_{3/2}\,\mu^2 \left(X + X^*\right) + 
\mathcal{O}\left(M_{\rm Pl}^{-2}\right)\,,
\label{eq:VXeff}
\end{align}
where $m_X$ denotes the effective sgoldstino mass induced at one loop.
According to our analysis in Sec.~\ref{subsec:spectrum}
and given the expression for the SUSY breaking parameter $\mu$ in
Eq.~\eqref{eq:WXSU}, it is now severely suppressed
by a large power of the ratio $\Lambda / M_{\rm Pl}$,
\begin{align}
m_X^2 \sim \frac{1}{16\pi^2} \left(\frac{m}{\mu}\right)^4 m^2 \sim
\frac{\kappa^4}{16\pi^2} \left(\frac{\Lambda}{M_{\rm Pl}}\right)^{4(N_c-2)} \Lambda^2 \,.
\end{align}
At the same time, the gravitino mass $m_{3/2}$ is also affected by
the suppression of the $\mu$ parameter,
\begin{align}
m_{3/2}^2 \simeq \frac{\mu^4}{3M_{\rm Pl}^2} \sim \kappa^2
\left(\frac{\Lambda}{M_{\rm Pl}}\right)^{2(N_c-1)} \Lambda^2 \,.
\end{align}
But for $N_c \geq 3$ colors, $m_{3/2}$ still ends up being larger
than the loop-induced effective mass $m_X$,
\begin{align}
\frac{m_X}{m_{3/2}} \sim \frac{\kappa}{4\pi}
\left(\frac{\Lambda}{M_{\rm Pl}}\right)^{N_c-3} \,.
\end{align}
In all $SU(N_c)$ theories with $N_c\geq 3$ colors, the scalar potential for
the sgoldstino field $X$ is therefore negatively curved to
leading order in $1/M_{\rm Pl}$, cf.\ Eq.~\eqref{eq:VXeff}.
This drives $X$ to a large VEV around the Planck scale,
$\left<\left|X\right|\right> \sim M_{\rm Pl}$, which eventually only becomes 
stabilized due to higher-dimensional SUGRA corrections in the scalar potential.
Such a large sgoldstino VEV is, however, fatal from the perspective of our
dynamical mechanism for the generation of an effective FI-term as well as
from the perspective of the entire $SU(N_c)$ DSB model.


First of all, it is inconsistent with our above derivation of
the $\xi$ parameter in Eq.~\eqref{eq:xiSU}, in which we assumed that
$\left<X\right> = \left<Z_0\right> = \left<\bar{Z}_0\right> = 0$.
But more than that, the fact that field $X$ is a linear combination
of $U(1)$ charge eigenstates implies that, now, the $U(1)$ flavor
symmetry is spontaneously broken at the Planck scale.
The $U(1)$ vector field $V$ thus acquires a mass of
$\mathcal{O}\left(M_{\rm Pl}\right)$, such that it
completely decouples from all low-energy physics.
Our low-energy effective SUGRA theory therefore no longer
features any $U(1)$ gauge interactions and there is no longer
any D-term potential that could possibly contain an effective FI-term.
Moreover, as far as the $SU(N_c)$ DSB model itself is concerned, we
are now facing a twofold Polonyi problem~\cite{Coughlan:1983ci}.
During the stage of cosmic inflation, both the real and the imaginary part
of the complex sgoldstino field are stabilized by means of a Hubble-induced
mass at field values that tend to be a distance of $\mathcal{O}\left(M_{\rm Pl}\right)$
away from the true vacuum at $\left<\left|X\right|\right> \sim M_{\rm Pl}$.
At the end of inflation, both real scalar fields hence begin to oscillate
around the true vacuum with very large initial amplitudes.
The subsequent decay of these oscillations then results in disastrous
amounts of entropy production, which threatens the successful generation of
the baryon asymmetry of the universe as well as primordial nucleosynthesis.
To avoid such catastrophic consequences, the $SU(N)_c$ DSB model needs to
be supplemented with a (dynamical) mechanism for the stabilization of 
the two real scalar DOFs contained in $X$.
The minimal setup  discussed above lacks such a mechanism.


In conclusion, we therefore find that our $SU(N_c)$ theories, while they represent
viable models of dynamical SUSY breaking in the rigid limit, are not suited for the
generation of an effective FI-term.
In addition, the SUSY breaking dynamics themselves are endangered and serious
cosmological problems arise once gravitational corrections are properly taken into account.
The origin of all these problems can be traced back to the suppression of the SUSY breaking scale,
$\mu \simeq \Lambda_{\rm SUSY}$, in the superpotential.
We expect that similar problems arise in every DSB model in which the nonzero
F-terms responsible for SUSY breaking originate from higher-dimensional operators
in the superpotential, i.e.\ in which the SUSY breaking scale is suppressed by
powers of the ratio $\Lambda / M_{\rm Pl}$.
A natural possibility to avoid the resultant theoretical and phenomenological
problems is then to focus on DSB models in which SUSY breaking
is driven by relevant or marginal operators in the Lagrangian.
Here, the $SP(N_c)$ models discussed in the present paper represent
a prime example of DSB models that fulfill exactly this requirement.


\section{Applications}
\label{sec:applications}


In the two previous sections, we have presented  and discussed in detail our dynamical
mechanism for the generation of an effective FI-term in vector-like models of dynamical
SUSY breaking.
After having thus completed our field-theoretic analysis, we shall now briefly comment
on possible phenomenological applications of our effective FI-terms in realistic models.
Here, we shall sketch in particular possible consequences of a nonvanishing FI-term
for the mediation of SUSY breaking to the visible sector, cf.\ Sec.~\ref{subsec:susybreaking},
as well as for cosmic inflation, cf.\ Sec.~\ref{subsec:inflation}. 


\subsection{Mediation of supersymmetry breaking to the visible sector}
\label{subsec:susybreaking}


Together with the choice of the mediation mechanism, the VEVs of the F- and D-terms
set the scale of the soft masses in the low-energy effective theory.
SUSY breaking will be mediated from the hidden to the visible sector
at least by gravitational interactions.
In this case, expanding the SUGRA scalar potential around the vacuum
leads to soft masses for the MSSM scalars,%
\footnote{Here, following the notation of Ref.~\cite{Wess:1992cp},
$D_i W = \partial W/ \partial \phi_i + M_{\rm Pl}^{-2} \, W \,\partial K/ \partial \phi_i$.}
\begin{align}
\begin{split}
V_F + V_D & = e^{K/M_\text{Pl}^2} \hspace{3.5pt}\left(K^{i\bar{\jmath}}\,D_i W \,D_{\bar{\jmath}}
 \overline W - 3\,\frac{\left|W\right|^2}{M_\text{Pl}^2}  + e^{-K/M_\text{Pl}^2} \,\frac{1}{2} D^2 \right) \\
 & = e^{K_0/M_\text{Pl}^2} \left(\frac{\left|W_0\right|^2}{M_{\rm Pl}^4}  \left|\phi_i\right|^2
   - \frac{1}{2} \frac{\left<D^2\right>}{M_{\rm Pl}^2}
   \left|\phi_i\right|^2 \right) + {\cal O}\big(\left|\phi_i\right|^4\big) + \dots \,,
 \end{split}
 \label{eq_VDVF}
\end{align}
The first term in parenthesis is the familiar contribution from F-term SUSY breaking,
\begin{align}
\left(m_0^F\right)^2 =  e^{K_0/M_\text{Pl}^2}\, \frac{\left|W_0\right|^2}{M_{\rm Pl}^4} = m_{3/2}^2 = 
\frac{\Lambda_{\rm SUSY}^4}{3 M_\text{Pl}^2} \,, \quad
\Lambda_{\rm SUSY}^4 = \left<|F_\text{tot}|^2\right> + \frac{1}{2} \left< D^2 \right> \,,
\end{align}
whereas the second term yields tachyonic contributions to the scalar masses,
\begin{align}
\left(m_0^D\right)^2 = - \frac{1}{2} \frac{\left<D^2\right>}{M_{\rm Pl}^2} \, e^{K_0/M_\text{Pl}^2} \,.
\label{eq:mssmneg}
\end{align}
Note that, in this setup, the MSSM scalars do not enter $V_D$.
The mass contribution in Eq.~\eqref{eq:mssmneg} is rather a direct
consequence of the requirement of a Minkowski vacuum, i.e.\ of requiring
the parenthesis in the first line of Eq.~\eqref{eq_VDVF} to vanish.
However, contrary to the case of pure D-term mediation, these tachyonic
contributions do not endanger the stability of the MSSM vacuum, since
$\left<\left|F\right|\right> > \left< D \right>$ in the parameter
range of interest.
The MSSM gauginos remain massless at this order
in $\left<\left|F\right|\right>/M_\text{Pl}$,
since the nonvanishing $U(1)$ gauge charges of
$Z_+$ and $Z_-$ do not allow a direct
coupling between the goldstino field
$X = \left(Z_+ + Z_-\right)/\sqrt{2}$ and the gauginos.
Moreover, the $R$ symmetry-conserving D-terms will
not contribute to ($R$-violating) gaugino masses.
In summary, in this minimal setup and in absence of any further SUSY
breaking sector, both the MSSM scalars and the gravitino,
cf.\ Eq.~\eqref{eq:m32LambdaSUSY},  obtain masses of
$\mathcal{O}\left(\Lambda^2/M_\text{Pl}\right)$.
For a value of the dynamical scale $\Lambda$ around the GUT scale,
$\Lambda_{\rm GUT} \sim 10^{16}\,\rm{GeV}$, this implies masses
of about $10^{14}$~GeV and hence a complete decoupling from the
standard model spectrum.
This can, however, be avoided by simply lowering
the value of the dynamical scale $\Lambda$ (which translates into a
smaller value of the strong coupling constant $g_s$ at the Planck scale).


The soft masses discussed above can be enhanced by adding
gauge mediation, i.e.\ by adding messenger particles which
transform under ${\cal G}_\text{GUT} \supset \text{SM}$ and which
couple to $X$ and $F_X$ (F-term gauge mediation) or are charged
under the $U(1)$ symmetry (D-term gauge mediation).
The former will yield positive contributions to the squared
masses of the scalars, which are suppressed by the messenger scale
instead of the Planck scale and which hence stabilize the scalars. 
As for the MSSM gauginos, only suppressed masses are
induced by F-term gauge mediation~\cite{Izawa:1997gs}.
Meanwhile, D-term gauge mediation  yields scalar
mass contributions of the following form~\cite{Nakayama:2007cf},
\begin{align}
 m_0^{D,\,\text{GM}} \sim \frac{\alpha}{4 \pi}
 \frac{\left< D^2 \right>}{\Lambda^3} \sim \frac{\alpha \, g^2}{4 \pi} \Lambda\,,
\end{align}
where $\alpha$ denotes the coupling strength of the
standard model gauge group at the GUT scale.
Unfortunately, the sign of this contribution is incalculable
due to the strong dynamics involved in the hidden sector.
Furthermore, $R$ symmetry-breaking gaugino condensation
in the strongly coupled sector can yield nonvanishing masses
for the MSSM gauginos,
\begin{align}
 m_{1/2}^{D,\,\text{GM}} \sim \frac{\alpha}{4 \pi} \frac{\left< D^4 \right>}{\Lambda^7}
 \sim \frac{\alpha \, g^4}{4 \pi} \Lambda \,.
\end{align}
For a further investigation on raising the gaugino
masses through D-term gauge mediation, cf.\ also
Ref.~\cite{Matos:2009xv}.
Finally, also when SUSY breaking is communicated to the
visible sector through anomaly mediation, the additional
D-term can help to stabilize the otherwise tachyonic scalars,
if the MSSM fields carry suitable charges under the $U(1)$
symmetry~\cite{Jack:2000cd}. 


In conclusion, we find that in phenomenologically viable
setups the sparticle mass scale is controlled
by the dynamical scale $\Lambda$, which is typically very
large, cf.\ Eq.~\eqref{eq:Lambdavalues},
far beyond the range of collider searches.
In order to obtain a lighter sparticle spectrum, we would have to take the gauge
coupling constant $g$ to extremely small values, which may be less well motivated
from a phenomenological point of view.
On the other hand, $g$ may be naturally suppressed for some specific physical reason.
A more thorough investigation of this question is left for future work.


\subsection{Cosmic inflation driven by a dynamical D-term}
\label{subsec:inflation}


In the previous section, we investigated the effect of the dynamically
generated FI-term on the mediation of SUSY breaking to the visible sector,
assuming that both $\left<\left|F\right|\right>$ and $\left< D \right>$
remain unchanged until today.
A further interesting situation arises when the D-term is responsible for a
phase of cosmic inflation, ending once the D-term is absorbed by the VEV of another
scalar field.
Ensuring a (nearly) vanishing cosmological constant in the true vacuum then
requires a cancellation among the contributions to the F-term potential,
$m_{3/2}^2 = \big<\left|F_\text{tot}\right|^2\big> /(3 M_\text{Pl}^2)$,
cf.\ Eq.~\eqref{eq:m32LambdaSUSY}.
Inflation is hence driven by the D-term potential only, even
though $\left< D \right> < \left<\left|F\right|\right>$.


D-term hybrid inflation is an attractive, simple realization of
cosmic inflation in SUSY~\cite{Binetruy:1996xj}.
The vacuum energy density driving inflation is provided by an (Abelian) FI-term.
At the same time, the slope of the inflationary potential arises from
radiative corrections due to perturbative Yukawa interactions of
the inflaton field encoded in the tree-level superpotential,
\begin{align}
 W_\text{inf} = \gamma \: \Phi_0 \, \Phi_+ \Phi_- \,.
 \label{eq_Winf}
\end{align}
Here, $\Phi_0$ contains the inflaton field $\phi$ and
the $\Phi_\pm$, carrying opposite charges under the
$U(1)$ symmetry associated with the FI-term, contain
the `waterfall' field, which is responsible for
ending inflation by absorbing the effective FI-term
in its VEV.
$\gamma$ denotes a Yukawa coupling.


D-term hybrid inflation typically features super-Planckian field values.
Hence, an embedding of the globally supersymmetric
model into an effective SUGRA framework is mandatory.
However, as discussed in Sec.~\ref{sec:introduction}, coupling a
constant FI-term to SUGRA poses serious difficulties and the
field-dependent FI-terms discussed in the literature so far
are generically fixed to the Planck scale.
This renders them useless for D-term hybrid inflation,
where the measurement of the amplitude of the scalar power
spectrum fixes the magnitude of the FI-term to a value
close to the GUT scale.
On the other hand, using the dynamically generated FI-term
described in the present paper, the implementation of
D-term hybrid inflation is straightforward.
As long as $\left< \Phi_\pm \right> = 0$, a large vacuum
energy density, $V_0 = \frac{1}{2}g^2\xi^2$,
is generated by the D-term, cf.\ Eq.~\eqref{eq:xi}.
Once the waterfall field absorbs $\xi$ in its VEV, the
D-term scalar potential vanishes and a flat Minkowski
vacuum (with SUSY broken by nonvanishing F-terms) is recovered.
In this context, successful D-term hybrid inflation
fixes the scale of SUSY breaking to
$\Lambda_\text{SUSY} \gtrsim \sqrt{\left|\xi\right|} \sim \Lambda_{\text{GUT}}$. 


This setup brings two important advantages over previous approaches.
First, the FI-term can be easily generated at the GUT scale,
as required by the data on the cosmic microwave background.
Second, since the local $U(1)$ symmetry under which the $\Phi_\pm$
are charged is broken by the meson VEVs from the outset,
no local cosmic strings are generated at the end of inflation.
In general, we expect $\Phi_\pm$ to couple to other fields
charged under the $U(1)$ symmetry via (higher-dimensional)
operators which are $U(1)$-invariant, but not not invariant
under $\Phi_\pm$ phase rotations,
e.g.\ $K \supset \Phi_- \Phi_- Z_+ Z_+ / M_\text{Pl}^2$.
In this case, the global symmetry associated with a
rotation of the VEV of the waterfall field in the
complex plane is explicitly broken, and no global
cosmic strings are formed.
If one forbids these terms by imposing an additional
symmetry, then the phase transition ending inflation may,
after all, produce topological defects associated with
the breaking of just this symmetry.
However, we stress that this is not the generic case
and hence the production of cosmic strings at the end
of inflation can be easily avoided.
This extends the parameter space of D-term hybrid
inflation to larger values of $\xi$ and hence smaller
values of the scalar spectral index $n_s$, i.e.\ into
a phenomenologically interesting regime, which is otherwise ruled out
by the observational bound on the tension of cosmic $U(1)$ strings~\cite{Battye:2010hg}.


A concrete realization of D-term hybrid inflation in SUGRA requires
the choice of a suitable K\"ahler potential to resolve the eta problem.
Here, an attractive possibility is to impose a shift symmetry along, say, 
the imaginary component $\text{Im}\{\phi\}$ of the inflaton field $\phi$, so that
the tree-level K\"ahler potential no longer depends on $\text{Im}\{\phi\}$,
but only on the real component $\text{Re}\{\phi\}$,
\begin{align}
\Phi_0 \rightarrow \Phi_0 + i\,\alpha\,  M_{\rm Pl} \,, \quad \alpha \in \mathbb{R} 
\quad \Rightarrow \quad K = \frac{1}{2}\big(\Phi_0 + \Phi_0^\dagger \big)^2 +
\Phi_+^\dagger \Phi_+ + \Phi_-^\dagger \Phi_- \,.
\end{align}
Such a choice of the K\"ahler potential then allows for D-term inflation taking place
in a chaotic regime, at field values high above the Planck scale,
$\text{Im}\{\phi\} \gg M_{\rm Pl}$.
The large inflaton field excursion $\Delta\phi$ in this scenario then results in a large tensor-to-scalar
ratio in accord with the large value recently claimed by the BICEP2 collaboration~\cite{Ade:2014xna}.
The idea of `chaotic D-term inflation` has been proposed and investigated
in Ref.~\cite{Domcke:2014xxx} and recently been applied in Ref.~\cite{Buchmuller:2014rfa}.
Alternatively, one may assume a K\"ahler potential of the no-scale type, which
leads to a small tensor-to-scalar ratio in accord with the bound on this
observable deduced from the PLANCK data~\cite{Ade:2013zuv}.
Such a scenario has, for instance, been studied in Ref.~\cite{Buchmuller:2012ex}.


\section{Conclusions and outlook}
\label{sec:conclusions}


The appearance of FI-terms in SUGRA is plagued by serious problems.
On the one hand, fundamental, genuine FI-terms are very difficult,
if not even impossible to realize in minimal SUGRA.
Field-dependent FI-terms, on the other hand, can arise in the low-energy
effective theory, e.g.\ via the Green-Schwarz mechanism in string theory.
But their magnitude is generically found to be restricted to values
around the Planck scale.
In this paper, we have proposed a dynamical mechanism
to generate a nonzero effective D-term which overcomes these limitations.


Our starting point is the IYIT model of dynamical SUSY breaking---a strongly
coupled supersymmetric $SP(N_c)$ gauge theory with $N_f = N_c + 1$ flavors.
In this DSB model, SUSY is spontaneously broken via the interplay of
the deformed moduli constraint and tree-level Yukawa interactions
that stabilize all flat directions in moduli space.
Gauging an Abelian subgroup of the total flavor symmetry, the nonvanishing
VEVs of the charged composite states at low energies then result (for generic values
of the Yukawa couplings) in a nonzero effective D-term.
This D-term can be interpreted as a field-dependent FI-term, the scale of which is
determined by the dynamical scale $\Lambda$.
As the scale $\Lambda$ can be freely varied over many orders of magnitude, our dynamical mechanism
hence allows for the generation of FI-terms at any scale between ${\cal O}(100)$~TeV and the
Planck scale.
Below the dynamical scale, but above the $U(1)$ vector boson mass scale
(which is suppressed by the small $U(1)$ gauge coupling constant),
the D-term acts as a genuine, constant FI-term.
Moreover, the scalar modulus field ensuring the gauge invariance of the FI-term
turns out to be automatically stabilized by the large IYIT F-term, rendering it
parametrically heavier than the gravitino.
The variable magnitude of our effective FI-term, the built-in dynamical
stabilization of the modulus field and the large hierarchy between the SUSY
breaking scale and the vector boson mass scale represent three major advantages of our
mechanism over its stringy alternative based on the Green-Schwarz mechanism.


A minimal realization of dynamical mechanism is accomplished in the $SP(1)$ version
of the IYIT model of dynamical SUSY breaking.
This minimal scenario already suffices to illustrate all key features
of our mechanism for the generation of an effective FI-term.
It can be straightforwardly extended to the more general scenario of
strongly interacting $SP(N_c)$ dynamics.
On the other hand, DSB models in which the SUSY breaking scale is
suppressed by some power of $\Lambda/M_\text{Pl}$ turn out to be problematic:
As we demonstrate for the case of $SU(N_c)$ dynamics, the $U(1)$ vector
field then acquires a very large mass, so that it completely decouples
from the low-energy physics.
At sub-Planckian energies, there is hence no longer an auxiliary $D$
field present in the theory, which precludes the possibility of having
an effective FI-term.
At the same time, the SUSY breaking dynamics themselves may be endangered,
once SUGRA corrections are properly taken into account.
This is an even more severe problem, which goes beyond the question of whether or not
it is possible to consistently generate an effective FI-term.


Dynamically generated D-terms open up a wide range of applications in SUSY model building.
SUSY breaking via nonvanishing F-terms is inherent in this setup and
its scale is set by the dynamical scale of the strong interactions; the
gravitino mass is in particular given by $m_{3/2} \sim \Lambda^2/M_\text{Pl}$.
In addition, nonvanishing D-terms contribute to the soft masses.
In particular in gauge mediation, this yields phenomenologically interesting
improvements over pure F-term SUSY breaking, as we have briefly reviewed in
Sec.~\ref{subsec:susybreaking}.
Alternatively, if $\Lambda$ is of order of the GUT scale, the dynamically generated
D-term can be the source of cosmic inflation.
Here, the simplest inflationary model featuring fields in the inflaton sector that
are charged under the $U(1)$ symmetry, D-term hybrid inflation, gives results
in accordance with current observations.
Remarkably enough, since the $U(1)$ gauge symmetry is spontaneously broken
from the outset, the cosmic string problem (which otherwise rules out the
simplest model) is now absent.


Open questions remain.
In this paper, we have focused on D-terms generated by gauging an Abelian flavor
symmetry present in a certain class of vector-like DSB models.
It remains to be investigated whether our mechanism can also be extended
to non-Abelian flavor symmetries as well as to alternative DSB models.
Furthermore, the mechanism presented here might be embedded into
conformal SUSY breaking models, which would promise the possibility
to combine a high SUSY breaking scale during inflation with a low SUSY
breaking scale in the true vacuum~\cite{Schmitz:2014xx}.
Moreover, we expect further possible applications in the context of model building
in supersymmetric gauge theories.
For example, given the possibility of taking $\Lambda$ to be of the
order of the GUT scale, it would be interesting to identify the $U(1)$
as part of a GUT group.
For example, an identification with $U(1)_{B-L}$, where $B$$-$$L$ denotes the difference
between baryon number $B$ and lepton number $L$, would allow to link the
scales of cosmic inflation, leptogenesis and SUSY breaking in an intriguing way.
However, in this case, the requirement that the effective FI-term not be
absorbed in the VEV of any standard model field adds additional constraints
and we leave a further investigation of this question to future work.


\subsubsection*{Acknowledgements}

The authors would like to thank F.~Br\"ummer, W.~Buchm\"uller, K.~Harigaya, M.~Hindmarsh,
R.~Kallosh, J.~Maltz, and C.~Wieck for helpful discussions and comments.
V.\,D.\ also gratefully acknowledges the hospitality of Kavli IPMU
during the initial stages of this project.
This work has been supported in part by the European Union through the FP7 Initial Training Network
INVISIBLES (Marie Curie Action PITAN-GA-2011-289442-INVISIBLES) (V.\,D.), by Grants-in-Aid
for Scientific Research from the Ministry of Education, Science, Sports, and
Culture (MEXT), Japan, No.\ 26104009 and No.\ 26287039 (T.\,T.\,Y.) as well as by the World Premier
International Research Center Initiative (WPI), MEXT, Japan (K.\,S.\ and T.\,T.\,Y.).



\end{document}